\documentstyle{mn}
 
\newif\ifAMStwofonts
\newcommand{\beq}{\begin{equation}}
\newcommand{\eeq}{\end{equation}}
 
\newcommand{\figl}[7]{
    \protect\centerline{
    \epsfxsize=#1\epsffile[#2 #3 #4 #5]{#6 angle=#7}}}
 
\def\lsim{\mathrel{\lower2.5pt\vbox{\lineskip=0pt\baselineskip=0pt
           \hbox{$<$}\hbox{$\sim$}}}}
\def\gsim{\mathrel{\lower2.5pt\vbox{\lineskip=0pt\baselineskip=0pt
           \hbox{$>$}\hbox{$\sim$}}}}

\ifoldfss
    \NewTextAlphabet{textbfit} {cmbxti10} {}
    \NewTextAlphabet{textbfss} {cmssbx10} {}
    \NewMathAlphabet{mathbfit} {cmbxti10} {} % for math mode
    \NewMathAlphabet{mathbfss} {cmssbx10} {} %  "   "    "
  \fi
  \ifAMStwofonts
    \ifCUPmtlplainloaded \else
      \NewSymbolFont{upmath} {eurm10}
      \NewSymbolFont{AMSa} {msam10}
      \NewMathSymbol{\upi}     {0}{upmath}{19}
      \NewMathSymbol{\umu}     {0}{upmath}{16}
      \NewMathSymbol{\upartial}{0}{upmath}{40}
      \NewMathSymbol{\leqslant}{3}{AMSa}{36}
      \NewMathSymbol{\geqslant}{3}{AMSa}{3E}

       \let\le=\leqslant
      \let\geq=\geqslant \let\ge=\geqslant
    \fi
  \fi
\ifnfssone
  \newmathalphabet{\mathit}
  \addtoversion{normal}{\mathit}{cmr}{m}{it}
  \addtoversion{bold}{\mathit}{cmr}{bx}{it}
  \newmathalphabet{\mathbfit} % math mode version of \textbfit{..}
  \addtoversion{normal}{\mathbfit}{cmr}{bx}{it}
  \addtoversion{bold}{\mathbfit}{cmr}{bx}{it}
  \newmathalphabet{\mathbfss} % math mode version of \textbfss{..}
  \addtoversion{normal}{\mathbfss}{cmss}{bx}{n}
  \addtoversion{bold}{\mathbfss}{cmss}{bx}{n}
  \ifAMStwofonts
    \ifCUPmtlplainloaded \else
      \UseAMStwoboldmath
      \makeatletter
      \new@mathgroup\upmath@group
      \define@mathgroup\mv@normal\upmath@group{eur}{m}{n}
      \define@mathgroup\mv@bold\upmath@group{eur}{b}{n}
      \edef\UPM{\hexnumber\upmath@group}
      \new@mathgroup\amsa@group
      \define@mathgroup\mv@normal\amsa@group{msa}{m}{n}
      \define@mathgroup\mv@bold\amsa@group{msa}{m}{n}
      \edef\AMSa{\hexnumber\amsa@group}
      \makeatother
      \mathchardef\upi="0\UPM19
      \mathchardef\umu="0\UPM16
      \mathchardef\upartial="0\UPM40
      \mathchardef\leqslant="3\AMSa36
      \mathchardef\geqslant="3\AMSa3E

       \let\le=\leqslant
      \let\geq=\geqslant \let\ge=\geqslant
    \fi
  \fi
\fi % End of NFSS release 1

\ifnfsstwo
  \DeclareMathAlphabet{\mathbfit}{OT1}{cmr}{bx}{it}
  \SetMathAlphabet\mathbfit{bold}{OT1}{cmr}{bx}{it}
  \DeclareMathAlphabet{\mathbfss}{OT1}{cmss}{bx}{n}
  \SetMathAlphabet\mathbfss{bold}{OT1}{cmss}{bx}{n}
  \ifAMStwofonts
    \ifCUPmtlplainloaded \else
      \DeclareSymbolFont{UPM}{U}{eur}{m}{n}
      \SetSymbolFont{UPM}{bold}{U}{eur}{b}{n}
      \DeclareSymbolFont{AMSa}{U}{msa}{m}{n}
      \DeclareMathSymbol{\upi}{0}{UPM}{"19}
      \DeclareMathSymbol{\umu}{0}{UPM}{"16}
      \DeclareMathSymbol{\upartial}{0}{UPM}{"40}
      \DeclareMathSymbol{\leqslant}{3}{AMSa}{"36}
      \DeclareMathSymbol{\geqslant}{3}{AMSa}{"3E}

       \let\le=\leqslant
      \let\geq=\geqslant \let\ge=\geqslant
    \fi
  \fi
\fi % End of NFSS release 2

\ifCUPmtlplainloaded \else
  \ifAMStwofonts \else % If no AMS fonts
    \def\upi{\pi}
    \def\umu{\mu}
    \def\upartial{\partial}
  \fi
\fi

\input epsf

\title[Breaking the ``Redshift Deadlock'' -- II]{Breaking the 
``Redshift Deadlock'' -- II: The redshift distribution for the submillimetre population of galaxies}
\author[Itziar Aretxaga {\it et al.}]
{Itziar Aretxaga$^{1}$, David H. Hughes$^{1}$, Edward L. Chapin$^{1}$, 
Enrique Gazta\~{n}aga$^{2,1}$,
\newauthor
James S. Dunlop$^{3}$, Rob J. Ivison$^{4}$ \\
$^{1}$Instituto Nacional de Astrof\'{\i}sica, \'Optica y Electr\'onica 
(INAOE), Aptdo. Postal 51 y 216, 72000 Puebla, Mexico \\
$^{2}$IEEC/CSIC, Edifici Nexus-201, c/ Gran Capit\'an 2-4, 
08034 Barcelona, Spain\\
$^{3}$Institute for Astronomy, University of Edinburgh, Blackford Hill, 
Edinburgh, EH9 3HJ, UK \\
$^{4}$Astronomy Technology Center, Royal Observatory,  Blackford Hill, 
Edinburgh, EH9 3HJ, UK \\
}

\pagerange{\pageref{firstpage}--\pageref{lastpage}}
\pubyear{2002}

\begin{document}

\maketitle 

\label{firstpage}

\begin{abstract}
Ground-based sub-mm and mm-wavelength blank-field
surveys have identified more than 100 sources, the majority of which
are believed to be dusty optically-obscured starburst galaxies.
Colours derived from various combinations of FIR,
submillimetre, millimetre, and radio fluxes provide the only currently
available means to determine the redshift distribution of this new
galaxy population.

In this paper we apply our Monte-Carlo photometric-redshift technique,
introduced in paper\,I (Hughes et al. 2002), to the multi-wavelength
data available for 77 galaxies selected at 850$\mu$m and 1.25\,mm.  
We calculate a probability distribution
for the redshift of each galaxy, which includes a
detailed treatment of the observational errors and uncertainties in
the evolutionary model.  The cumulative redshift distribution of the
submillimetre galaxy population 
that we present in this paper, based on 
50 galaxies found in wide-area SCUBA surveys,
is asymmetric, and 
broader than those published elsewhere, with a significant high-$z$
tail for some of the evolutionary models considered.  
Approximately 40 to 90 per cent of the sub-mm population is expected to
have redshifts in the interval $2 \le z \le 4$.  Whilst this result is
completely consistent with earlier estimates for the sub-mm galaxy
population, we also show that the colours of many ($\lsim 50$ per cent )
individual sub-mm sources,  detected only at
850$\mu$m with non-detections at other wavelengths, 
 are consistent with those of starburst
galaxies that lie at extreme redshifts, $z > 4$.  Spectroscopic
confirmation of the redshifts, through the
detection of rest-frame FIR--mm wavelength molecular transition-lines,
will ultimately calibrate the accuracy of this technique. We use the
redshift probability distribution of HDF850.1 to illustrate the
ability of the method to guide the choice of possible frequency
tunings on the broad-band spectroscopic receivers that equip the large
aperture single-dish mm and cm-wavelength telescopes.
\end{abstract}

\begin{keywords}
submillimetre, millimetre, cosmology, galaxy evolution, star-formation
\end{keywords}

\section{Introduction}
The lack of robust redshift information is one of the outstanding problems in
understanding the nature of the presumed high-redshift galaxies
identified in blank-field sub-mm and mm wavelength surveys. 

In paper\,I of this series \cite{hughes02} we described Monte Carlo
simulations of 250--500$\mu$m surveys from the balloon-borne telescope
BLAST \cite{devlin01} and the SPIRE instrument on the Herschel
satellite, as well as longer-wavelength 
ground-based 850$\mu$m surveys with SCUBA. These
simulations included a detailed treatment of the observational and
evolutionary model-dependent errors.  The sub-mm colours of galaxies
derived from these mock surveys demonstrated that it will be
possible to derive photometric redshifts, with a conservative r.m.s.
accuracy of $\Delta z \sim \pm 0.5$, for thousands of
optically-obscured galaxies.  Thus we can look forward to 
breaking the redshift deadlock which, at the moment, prevents a
reliable estimate of the evolutionary history of a population that
contributes a significant fraction ($\gsim 50$ per cent for $S_{850\mu 
{\rm m}}>2$~mJy) to the sub-mm background
(Hughes et al. 1998, Smail et al. 2002).

In this second paper we apply the same Monte Carlo photometric-redshift 
technique to the existing FIR--radio multi-wavelength data
for 77 sources  first identified in
blank-field 850$\mu$m (SCUBA) and 1.2\,mm (MAMBO) surveys.
We calculate the redshift
probability distributions of the individual sub-mm and mm galaxies, taking
into account observational and model-dependent uncertainties,
and thus provide a measurement of the cumulative redshift
distribution for the blank-field sub-mm galaxy population. This
redshift distribution is completely insensitive to ambiguities in the 
identification of the  optical counterparts.

The structure of the paper is as follows: section~2 gives a brief
description of the Monte Carlo simulations, the method to derive
photometric redshifts from the radio--sub-mm--FIR colours of galaxies 
drawn from
the mock catalogues, and the caveats and advantages of this technique
over other redshift indicators; section~3 presents individual redshift
probability distributions for blank-field sub-mm galaxies.  In
addition, we include a brief discussion of the photometric redshifts
for radio and FIR-selected galaxies subsequently detected at
850$\mu$m. Furthermore, we determine the accuracy of our method 
through the comparison of the photometric redshifts for those 8 sub-mm
sources that have published rest-frame optical spectroscopic redshifts.
Finally, in section 3 we present the cumulative redshift
distribution of the blank-field sub-mm population. Section~4 discusses
the likelihood of a radio detection of previously detected blank-field
sub-mm sources, based only on the known dispersion of the
850$\mu$m/1.4\,GHz colour in starburst galaxies, radio-quiet AGN and
ULIRGs,
which are believed to be representative
of the high-$z$ sub-mm galaxy population.  
Section~5 describes how the
determination of redshift probability distributions for individual
sub-mm galaxies is motivating the efforts to provide spectroscopic
confirmation of the redshifts at mm and cm wavelengths. Ultimately,
these spectroscopic redshifts will accurately calibrate the
photometric redshifts derived from rest-frame FIR--radio-wavelength
data. Appendix~A contains the complete catalogue of spectral energy 
distributions (SEDs) and redshift
distributions derived in this paper.

\section{Photometric redshift estimation technique}

The Monte Carlo simulations described in this paper produce the
redshift probability distribution for an individual galaxy.  The
advantage of this technique, compared to the popular
maximum-likelihood methods, is that it provides more information than
a simple estimate of the first and second moments of the redshift
distributions. A detailed description of the technique can be found in
paper\,I; we only offer a summary here.  Hereafter we will refer
to a galaxy detected in a SCUBA or MAMBO survey as a {\em sub-mm}
galaxy, and a sub-mm galaxy generated in the Monte Carlo simulations
as a {\em mock} galaxy.

We choose an evolutionary model for the $60\mu$m luminosity function
that fits the observed 850$\mu$m number-counts. Under this model, and
assuming a sub-mm survey of $\sim$10 deg$^{2}$, we generate a
catalogue of 60$\mu$m luminosities and redshifts.  Randomly selected
template SEDs are drawn from a library of local starbursts, ULIRGs and
AGN, to provide FIR--radio fluxes, and hence colours, for this mock
catalogue.  The fluxes of the mock galaxies include both photometric
and calibration errors, consistent with the quality of the
observational data for the sub-mm galaxy detected in a particular
survey. We reject from the catalogue
those mock galaxies that do not respect the detection thresholds and
upper-limits of the particular sub-mm galaxy under analysis.  
In this paper the redshift
probability distribution of a sub-mm galaxy is calculated as the
normalized distribution of the redshifts of the mock galaxies in the
reduced catalogue, weighted by the likelihood of identifying the
colours and fluxes of each mock galaxy with those of the sub-mm galaxy
in question.  Thus, we also take into account as much
prior information on the sub-mm
galaxy population (and their uncertainties) as possible. For example, 
we consider uncertainties in the
sub-mm galaxy number counts, their favoured evolutionary models, the
luminosity function of dusty galaxies, lensing by a foreground
magnifying cluster, etc.  In the case that the sub-mm galaxy 
is identified as a lensed source, then the mock catalogue is also amplified
by the appropriate factor, allowing intrinsically fainter objects to
be introduced into the reduced catalogue, using the same observational
measurement errors. 

We adopt a flat, $\Lambda$-dominated cosmological model
($H_0=67$~km\,s$^{-1}$\,Mpc$^{-1}$, $\Omega_{\rm M}=0.3$,
$\Omega_{\Lambda}=0.7$).  To quantify the sensitivity of the
individual redshift distributions on the assumed evolutionary history
of the sub-mm galaxy population, we consider six different models
that are able to reproduce the observed $850\mu$m number-counts within
the uncertainties.
We will refer to them by the following codes:
\begin{description}
\item[{\bf le1:}] luminosity evolution of the $60\mu$m luminosity function 
$\Phi[L,z=0]$ (Saunders et al. 1990) as $(1+z)^{3.2}$ for $0 < z < 2.2$. 
Constant evolution is imposed between $z=2.2 - 10$.
No galaxies beyond $z=10$, $\Phi[L,z > 10] = 0$. %17 
\item[{\bf le1L13:}] same evolutionary description as in le1, but imposing a 
luminosity cut-off above $10^{13}L_{\odot}$.     %17f
\item[{\bf le2:}] luminosity evolution of $\Phi[L,0]$ as $(1+z)^{3}$ for 
$0 < z < 2.3$.  Constant evolution is imposed between $z=2.3 - 6$.  
$\Phi[L,z > 6] = 0$.                             %15
\item[{\bf le2L13:}] same evolutionary description as in le2, but imposing a 
luminosity cut-off above $10^{13}L_{\odot}$.     %15f

\item[{\bf lde1}] {\em  luminosity evolution of $\Phi[L,0]$  as $(1+z)^{1.9}$
  and density evolution as $(1+z)^{2.5}$ for 
$0 < z < 3.5$; luminosity and density evolution as $(1+z)^{-3.0}$ 
for $3.5 \le z \le 7$} %22
\item[{\bf lde2}] {\em luminosity evolution of $\Phi[L,0]$  as $(1+z)^{3.5}$
  and density evolution as $(1+z)^{3}$ for 
$0 < z < 2.0$; no further luminosity evolution, and density evolution 
as $(1+z)^{-2}$ for $2 \le z \le 6$ %24}
}
\end{description}

\noindent
The models used in 
this analysis provide a wide range of the many possible evolutionary scenarios
that reproduce the sub-mm to mm
number counts.
These models will allow us to test 
the robustness of the
photometric redshifts that we will derive in  the next section.

An extended library of SEDs (compared to paper I) is used in the
Monte Carlo simulations. The library contains 20 starbursts, ULIRGs
and AGN that have been either
extensively mapped at FIR--radio wavelengths, or galaxies that are
sufficiently distant that single-beam photometric observations measure
the total flux:
M\,82, NGC\,3227, NGC\,2992, NGC\,4151,
NGC\,7469, NGC\,7771, NGC\,1614, I\,Zw\,1, Mkn\,231, Arp\,220,
Mkn\,273, NGC\,6240, UGC\,5101, IRAS\,10214+4724, IRAS\,05189$-$2524,
IRAS\,08572+3915, IRAS\,15250+3609, IRAS\,12112+0305,
IRAS\,14348$-$1447 and Cloverleaf. 
This selection has the important consequence of
reducing any dispersion in the SEDs due to aperture effects. A
continuum SED that includes
synchrotron and free-free
emission  in the radio to mm regime,
and thermal emission from dust grains in the mm-to FIR regime,  is
fitted to the observational data (Chapin et al. 2003). 
The luminosities and temperatures of the dust that dominates the
rest-frame FIR emission are evenly distributed
in 
the ranges $9.0 \lsim \log L_{\rm
FIR}/L_{\odot} \lsim 12.3$ and $25<T/K<65$.

Half of the galaxies in the SED library used in the derivation of
photometric redshifts in this paper are in common with those
considered by Yun \& Carilli (2002), hereafter YC02.  The temperature
range covered by our sample of galaxies ($25-65$K, median 41K) is
broader than that of YC02 due to the increased fraction of cooler
sources, and is, for the same reason, colder than those of
well-studied ULIRGs ($46-77$K, median 60K, Klaas et al. 2001).
Despite the similar temperatures of our sample to late-type galaxies
selected at 60$\mu$m ($25-55$K, median 35K), drawn from the SLUGS
catalogue of Dunne et al. (2000), the shapes of the SEDs in our
library differ significantly from the local galaxies in the SLUGS.
This illustrates an important point: it is the shapes of the SEDs, and
not the derived dust temperatures, that influence the estimated
photometric redshifts.  Given the limited FIR--sub-mm data, there is a
degeneracy between the dust temperatures, grain emissivity index
($\beta$), source-size ($\Omega$) and optical depth ($\tau_\nu$).
This makes it possible to fit a single observed SED with a broad range
of temperature, by tuning the choice of $\beta, \Omega$ and
$\tau_\nu$ (e.g. Hughes et al 1993).

It remains unclear whether the lower-luminosity, local galaxies in the
SLUGS are closely related
to the blank-field SCUBA population which have rest-frame FIR
luminosities $> 10^{12} L_{\odot}$, assuming the sub-mm galaxies lie
at $z > 1$.  The claim that cold SEDs match the number-counts in
hydrodynamical cosmological simulations ({\it e.g.} Fardal et
al. 2002) should not be a basis for favouring cold SEDs, since a
different evolutionary model with warmer SEDs can fit the counts
equally well.

The strongest caveat in any photometric redshift analysis is the
validity of the assumption that the SED templates of local starbursts,
ULIRGs and AGN are good analogs of the high-redshift sub-mm galaxy
population. It is not until future rest-frame FIR--sub-mm observations
from sensitive balloon-borne or satellite experiments (paper\,I)
accurately measure the variety of SEDs present in the sub-mm galaxy
population, and spectroscopic redshifts for these are available, 
that it will possible to improve the error estimates
presented in this paper.

The technique adopted in this paper differs philosophically from those
developed in other photometric-redshift studies.  Our redshift
estimates do not rely on the match of the observed data to a single
template SED, and our errors are not drawn from the departure of other
local SEDs from the adopted standard template, which has become the
common practice (Carilli \& Yun 1999, 2000, hereafter referred to as
CY99, CY00; Gear et al. 2000; YC02).  Instead, we allow the test
galaxies to be matched against the whole variety of template SEDs that
can be detected at any given redshift.  This point should be
emphasized, since although we assign our template SEDs at random, not
all SED shapes (scaled to the same FIR luminosity) are detected with
equal probability at any given redshift and wavelength.  The redshift
distributions derived in this paper are often asymmetric, and broader
than those inferred from the single SED template analysis. Despite the
efforts of some authors to include realistic error-bars in their
single-template estimates, we consider these to have been
underestimated: error bars derived from 68\% of the galaxies in a
template catalogue, is not equivalent to the 68\% dispersion in the
fluxes that the whole catalogue can create, since some of the outliers
contribute very significantly to the departures from the mean.

Our redshift estimation method also has the advantage of including a
common-sense evaluation of whether a source at a given redshift is
simultaneously consistent with the observed colours and fluxes.  The
probability of generating a certain luminosity at a certain redshift
({\it i.e.} flux) is introduced by the luminosity function and the
evolutionary model adopted. As shown below, when sub-mm galaxies with
detections at more than two wavelengths are considered, the results
are quite insensitive to the differences in the evolutionary models 
adopted to generate
the mock catalogue. 
The consideration of reasonable evolutionary models, however, allow
plausible colour-based redshift solutions to be excluded.

\section{Monte Carlo based photometric redshifts}

The available data-set for the sub-mm population of galaxies is
usually restricted to a low-S/N detection at 850$\mu$m and an
upper-limit at 450$\mu$m (from SCUBA surveys), or a 1.2\,mm detection
(from MAMBO surveys or an IRAM PdB follow-up observation), 
with, perhaps, radio observations at
1.4, 5 or 8.6\,GHz.  Although a number of the SCUBA surveys have been
conducted in fields previously observed by the Infrared Space
Observatory (ISO)  with
the ISOPHOT camera at 170 and 90$\mu$m, 
{\it e.g.} HDF \cite{hughes98,borys02}, ELAIS N2
 and the Lockman Hole \cite{scott01,fox01}, the ISO measurements are
generally too insensitive to provide any additional constraint on the
photometric redshifts. 

The method described in section\,2 produces mock-catalogues that
replicate the sensitivities and calibration accuracies of the original
surveys in which each sub-mm galaxy was detected. The colours and
fluxes of the mock galaxies provide diagnostic diagrams to illustrate the
derivation of the most probable redshift: 
colour-colour-redshift ($C-C-z$) diagrams,
when more than 3-band detections are available; colour-flux-redshift
($C-f-z$) diagrams, when just two bands are available; and
flux-redshift ($f-z$) diagrams; when just one band detection is
available. From these diagrams we derive the corresponding redshift
probability distribution ($P-z$) for each sub-mm galaxy.
The distributions for a selection of  sub-mm galaxies 
considered in this paper are shown in Appendix\,A.

\subsection{Specific examples: LH850.1 and BCR11}

LH850.1 is the brightest 850$\mu$m source identified in the wide-area
UK 8~mJy SCUBA survey of the Lockman Hole \cite{scott01}, and has one of the
most complete FIR--radio SEDs \cite{lutz01}.  BCR11, named after
source\,11 in the sub-mm catalogue of Barger, Cowie \& Richards (2000), is a low S/N,
but nonetheless representative, 850$\mu$m source detected in a
follow-up SCUBA photometry survey of the micro-Jansky 1.4\,GHz radio
sources in the Hubble Deep Field \cite{richards00}.

Even though BCR11 was not first identified in
a blank-field sub-mm survey, and hence is not included in the combined
redshift distribution of the sub-mm selected galaxy population, 
it provides a useful
source to discuss. The sub-mm photometry observations of BCR11 at the
position of a known radio source \cite{richards00} guarantees that
we have the correct association of a radio and sub-mm source, and hence the
correct SED to analyse. BCR11 allows us to confidently derive the
photometric redshift and accuracy that we can expect from only the
850$\mu$m/1.4\,GHz colour, which is commonly used to measure redshifts of
blank-field sub-mm galaxies with follow-up radio observations (CY99, CY00).

\begin{figure}
    \vspace*{4cm}
    \hspace*{10cm}
    \figl{12cm}{26}{390}{573}{737}{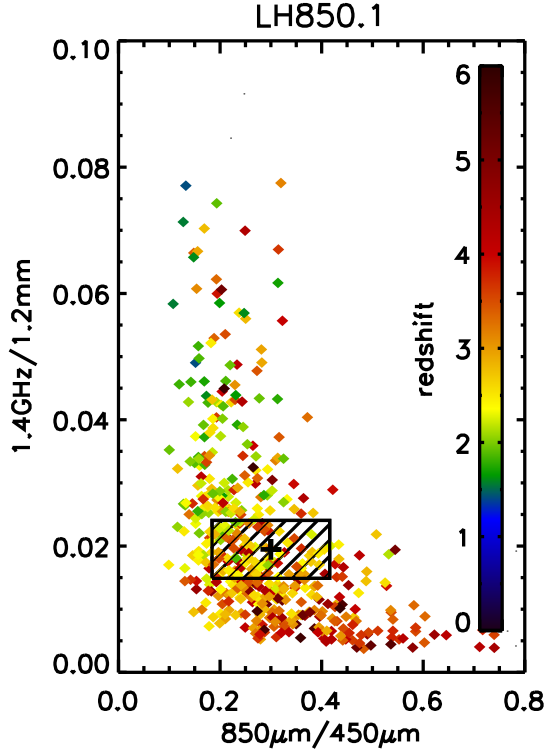}{90} 
%    \figl{12cm}{26}{26}{573}{737}{photz1_lh.ps}{90} 

    \caption{Colour-colour-redshift ($C-C-z$) plot for LH850.1 (Lutz
 et al. 2001, Scott
 et al. 2002). 
The flux ratios of the mock galaxies, generated under the evolutionary
 model le2 (section 2), are represented as diamonds, and their
 redshifts are coded in colour according to the scale shown to the
 right of the panel (colours can be seen in the electronic version of
 the paper).  
 The cross represents the
 measured colours of LH850.1, and the dashed box shows the 1$\sigma$
 uncertainty in each colour.}
\label{fig:LH_ccz}
\end{figure}

\begin{figure}
%    \vspace*{-2.5cm}
%\hspace*{8cm}
%    \vspace*{-2.5cm}
%\hspace*{16cm}
    \vspace*{-2.cm}
    \hspace*{8cm}
    \figl{7cm}{26}{26}{573}{737}{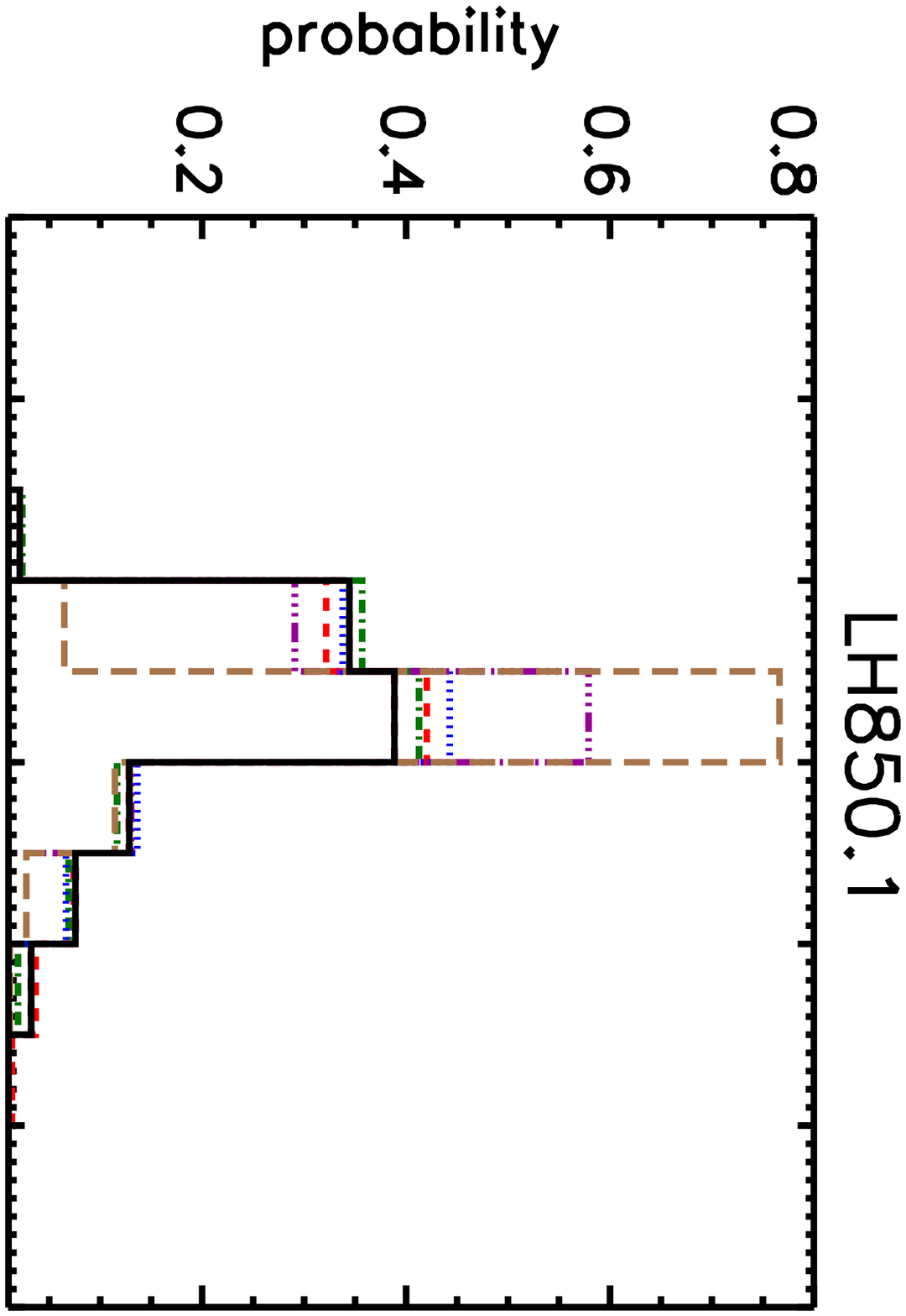}{90} 

    \vspace*{-4.6cm}
    \hspace*{8cm}
    \figl{7cm}{26}{26}{573}{737}{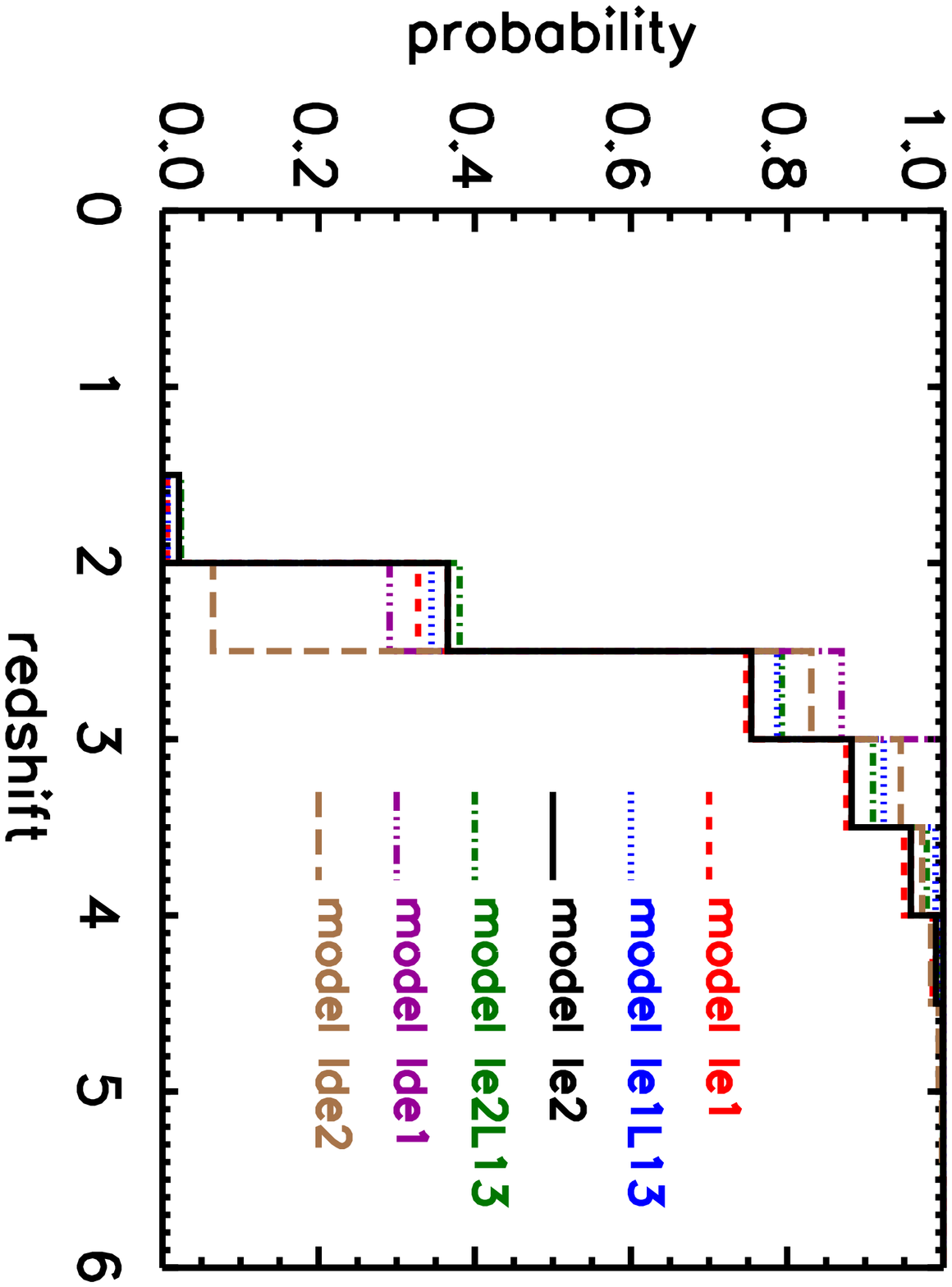}{90} 
\caption{Discrete and cumulative 
redshift probability distributions of LH850.1. The six estimates,
plotted with lines of different style, correspond to the six
evolutionary models introduced in section 2. }
\label{fig:LH_pz}
\end{figure}

\begin{figure}
    \vspace*{-2.5cm}
\hspace*{8cm}
    \figl{7cm}{26}{26}{573}{737}{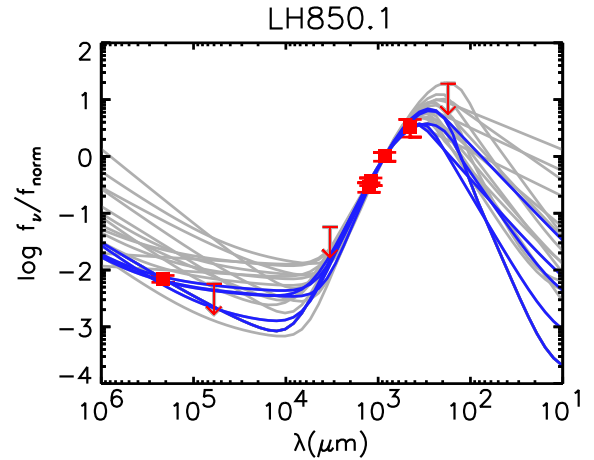}{90} 
    \caption{The observed SED of LH850.1 
    normalised to the flux density at 850$\mu$m is shown as
    squares and
    arrows. The arrows indicate 
    3$\sigma$
    upper limits. The squares denote detection at a level 
\hbox{$\geq 3\sigma$}, 
    with 1$\sigma$ error bars. 
    The template SEDs (lines) are redshifted to $z=2.6$, the mode of
    the redshift probability distribution of most models considered
    (see table 1). The template SEDs
{\em at this redshift} compatible 
within $3\sigma$ error bars with the SED of LH850.1 are displayed as
    darker lines (blue in the electronic version).
}
\label{fig:LH_sed}
\end{figure}

\begin{figure}
    \vspace*{4cm}
    \hspace*{10cm}
    \figl{12cm}{26}{390}{573}{737}{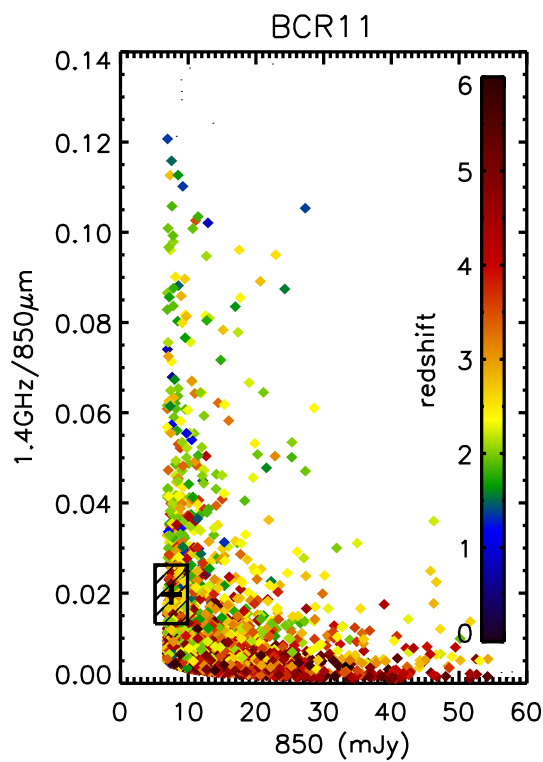}{90} 
%    \figl{12cm}{26}{26}{573}{737}{photz1_lh.ps}{90} 

    \caption{Colour-flux-redshift ($C-f-z$) plot for BCR11 (Barger et al. 2000). 
The description of the
symbols is as in Fig.~\ref{fig:LH_ccz}}
\label{fig:B11_ccz}
\end{figure}

\begin{figure}
    \vspace*{-2.cm}
    \hspace*{8cm}
    \figl{7cm}{26}{26}{573}{737}{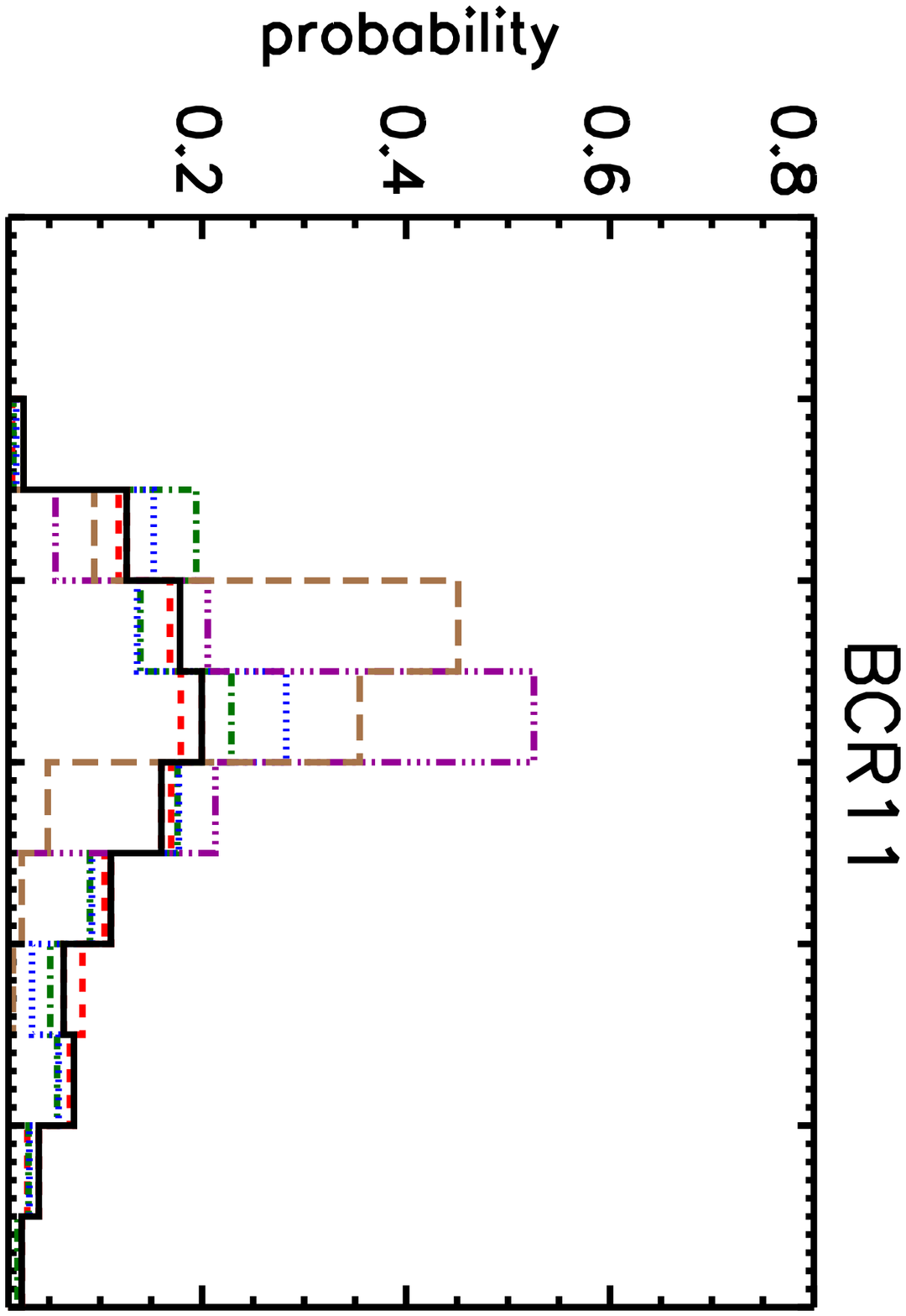}{90} 

    \vspace*{-4.6cm}
    \hspace*{8cm}
    \figl{7cm}{26}{26}{573}{737}{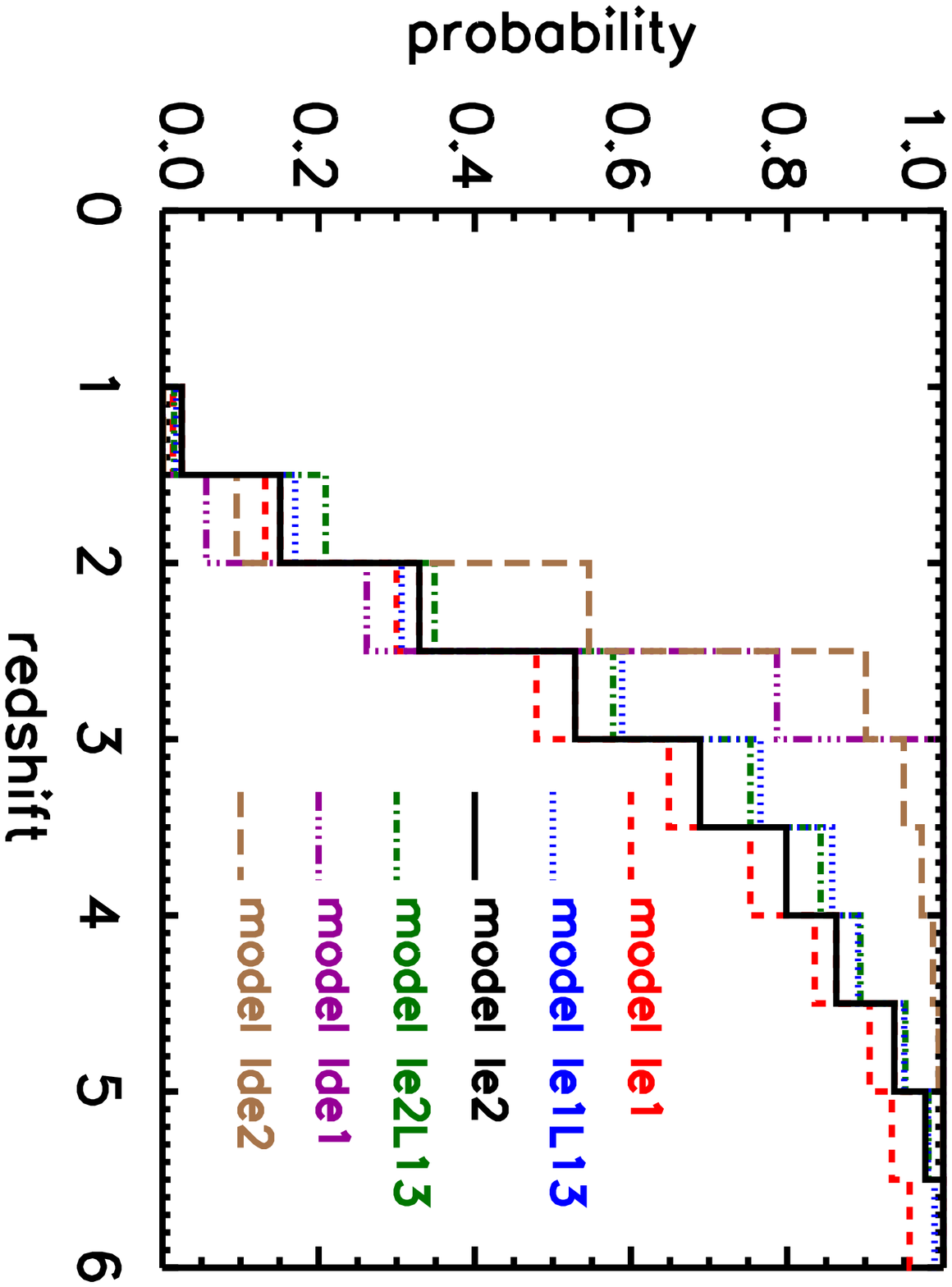}{90} 
\caption{Redshift probability distributions for BCR11. The description
of the plot is as in Fig.~\ref{fig:LH_pz} }
\label{fig:B11_pz}
\end{figure}

\begin{figure}
    \vspace*{-2.5cm}
\hspace*{8cm}
    \figl{7cm}{26}{26}{573}{737}{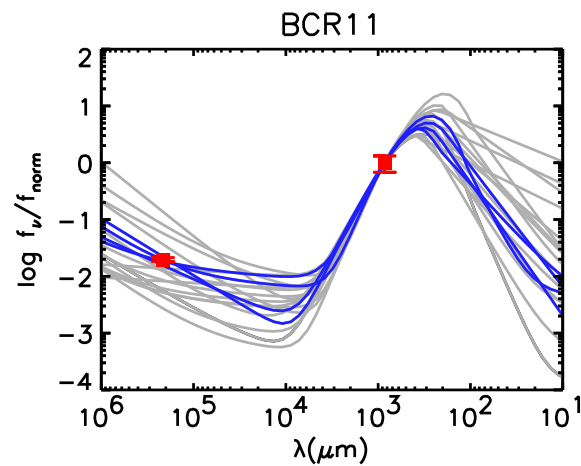}{90} 
\caption{Comparison of the observed SED of BCR11 with those of
template galaxies, redshifted to $z=2.7$ (table 2). 
The description of the plot is as in
Fig.~\ref{fig:LH_sed}. Note that the selection of SEDs to represent
BCR11 is different from that in LH850.1.}
\label{fig:B11_sed}
\end{figure}

Figs.~\ref{fig:LH_ccz},\ref{fig:LH_pz} and \ref{fig:LH_sed}, respectively,
show the $C-C-z$ and $P-z$ distributions for the multi-wavelength data
of LH850.1, and the comparison of the observed SED and template SEDs
(redshifted to the mode of the $P-z$
distribution). Figs.~\ref{fig:B11_ccz},\ref{fig:B11_pz} and \ref{fig:B11_sed}
show the same distributions for BCR11.  In both examples
(Figs.~\ref{fig:LH_pz} and \ref{fig:B11_pz})  the peak of the
  redshift probability distribution
depends little on the details of the evolutionary model used to
compute the mock catalogues. In both cases, the models that  
invoke pure luminosity evolution (le1, le1L13, le2, le2L13)
have more extended wings, since
they provide a significant population of sources at redshifts beyond 3 that can reproduce
the colours of the observed galaxies. On the other hand, the
combined luminosity and density evolution models (led1, led2) 
have a very strong
decline in the density of sources at redshifts beyond 2 and 3.5, and thus,  the
corresponding redshift distributions are much more concentrated to
values between 2 and 3. Model led1 is the  most extreme of these two
evolutionary models,
since the very strong negative luminosity and density evolution
does not allow any source beyond redshift $z=3.5$ to be detected at
any reasonable depth.
The concentration effect is more dramatic on BCR11,
where the colour constraints are  the weakest, and the wings of the
redshift distribution are strongly dependent on the prior information
on  the galaxy population that has been used. The sharpness of this
distribution is, thus, due to the prior, and not to the colour
constraints. In contrast, the sharpness of the distribution of LH850.1
is intrinsic to the well determined 
colours of  the source, and is almost independent
of the prior.

The comparison of LH850.1 and BCR11  
demonstrates the improved redshift
accuracy that, 
in general, one can expect for those sources with the greatest
combination of deep multi-wavelength data. In the case
of BCR11, it also illustrates the difficulty of constraining
individual redshifts
with observations
 at 1.4\,GHz and 850$\mu$m only.  The redshift probability
distributions of both sources have very similar modes
($z=2.6-2.8$), yet for the less constraining evolutionary models
(le1, le2, le1L13, le2L13)
there remains 40--50\% and 25--35\% probabilities that
BCR11 has a redshift $>3$ and $>4$ respectively.  The
corresponding probabilities for LH850.1 are 25-30\% and 5\% respectively.
For models with a strong decline in the number of sources with
redshift (led1, led2), the high-$z$ tails are not produced. 
The sharpness of
the derived redshift distributions is due to the lack
of bright sources at high redshift, and not by the colours of the galaxy.

\begin{figure*}
    \vspace*{-2.0cm}
    \hbox{ \hspace{3cm} \figl{6cm}{26}{26}{573}{737}{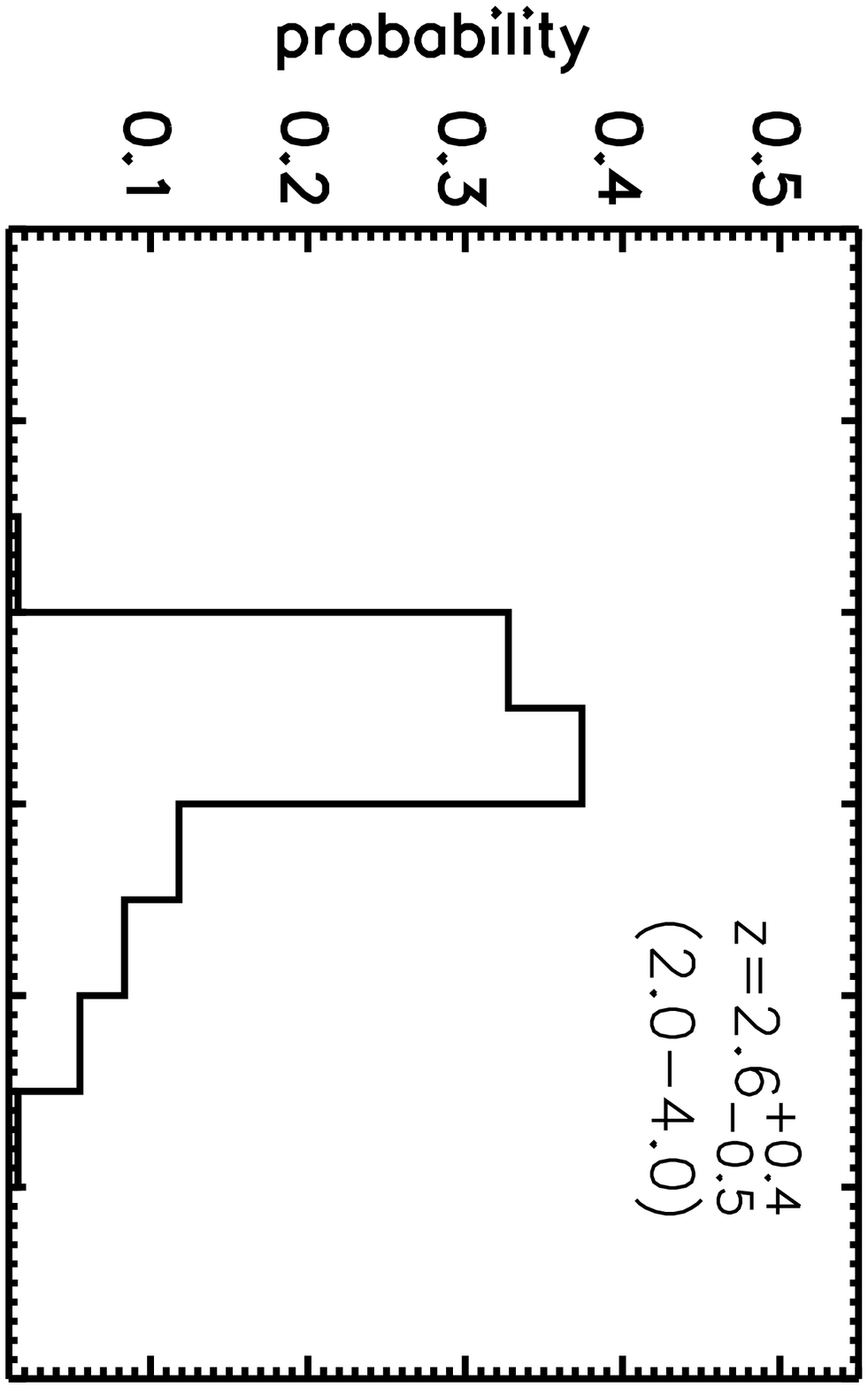}{90} 
           \hspace{-10cm} \figl{6cm}{26}{26}{573}{737}{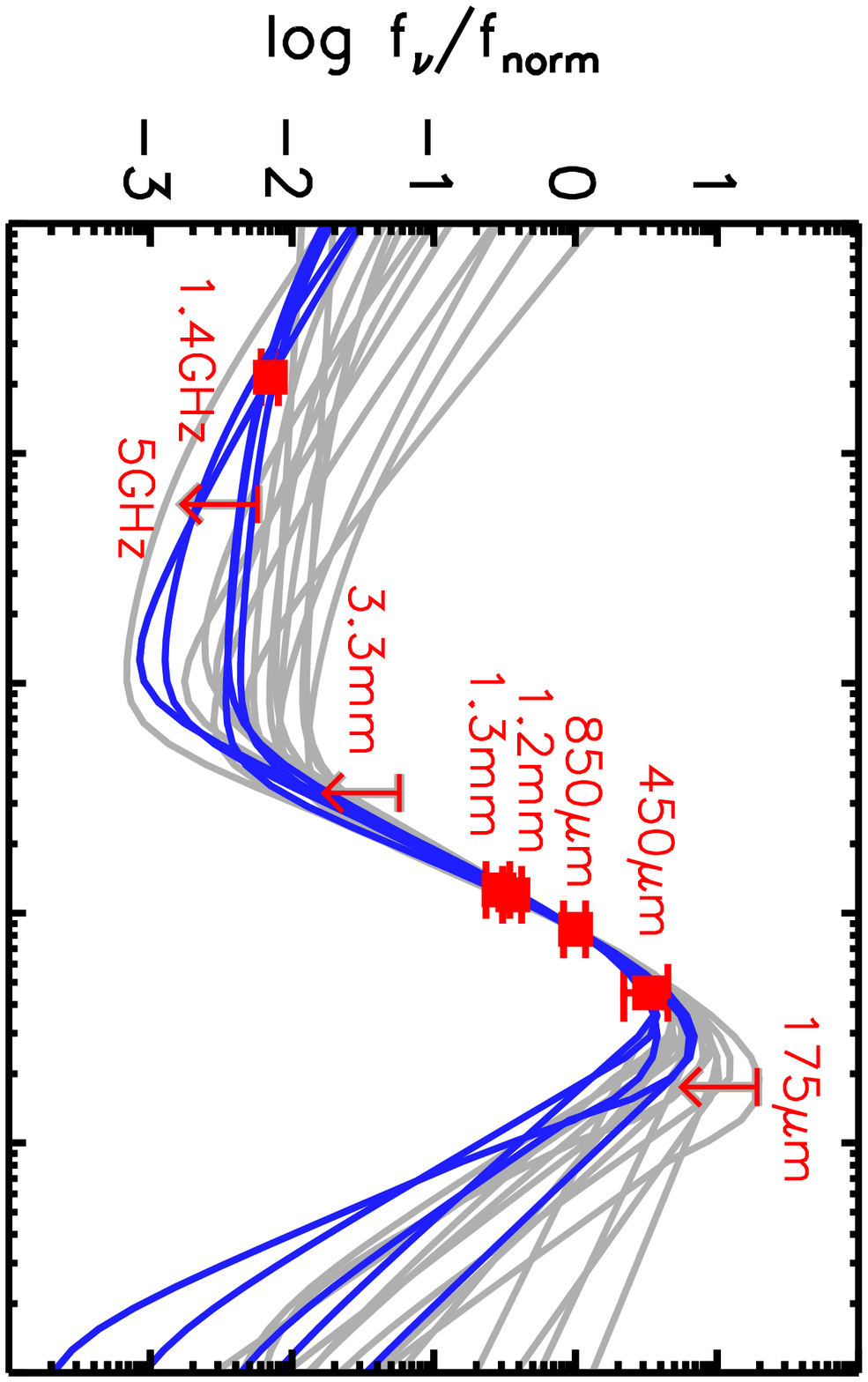}{90} }

    \vspace*{-3.95cm}
    \hbox{ \hspace{3cm} \figl{6cm}{26}{26}{573}{737}{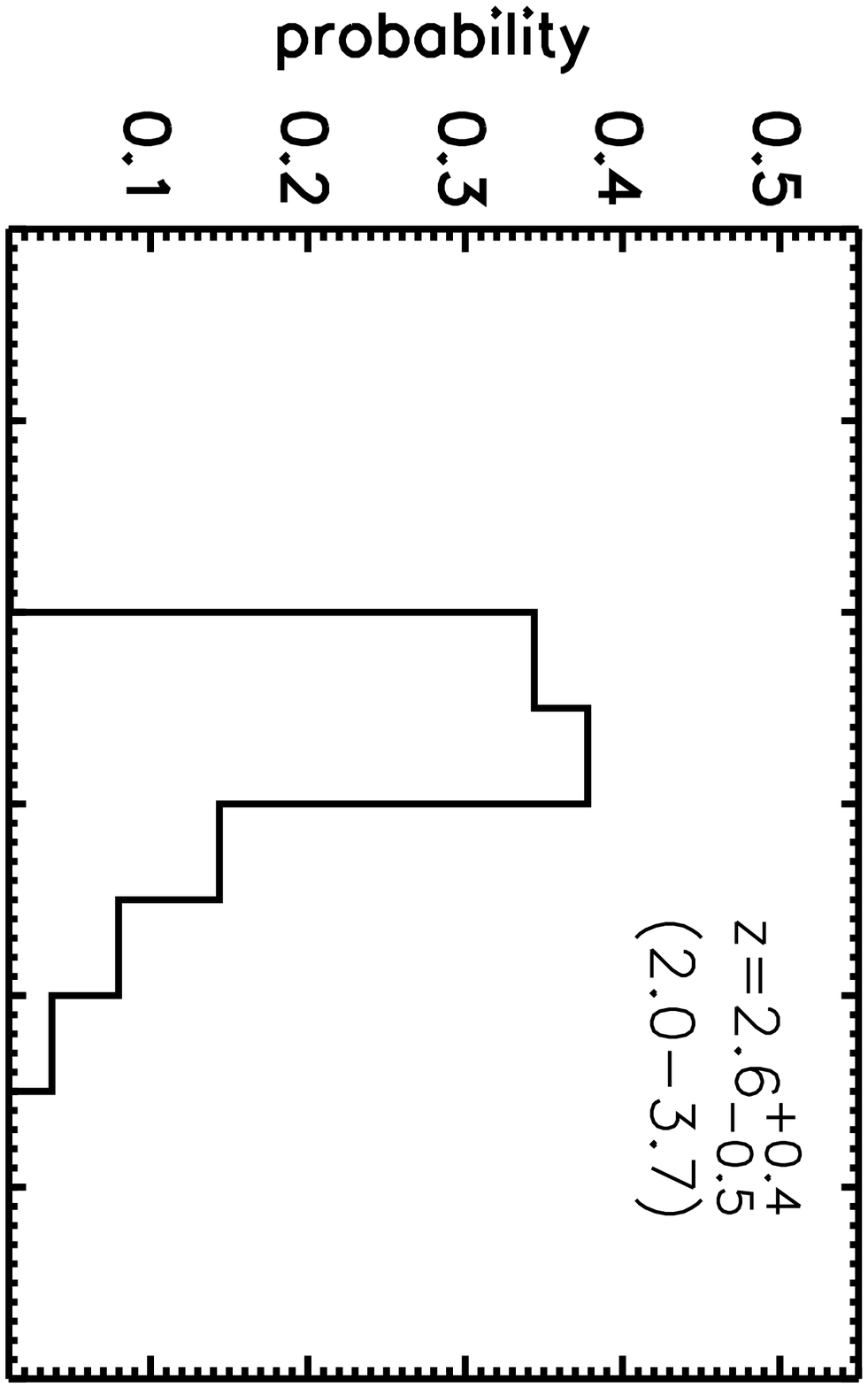}{90} 
           \hspace{-10cm} \figl{6cm}{26}{26}{573}{737}{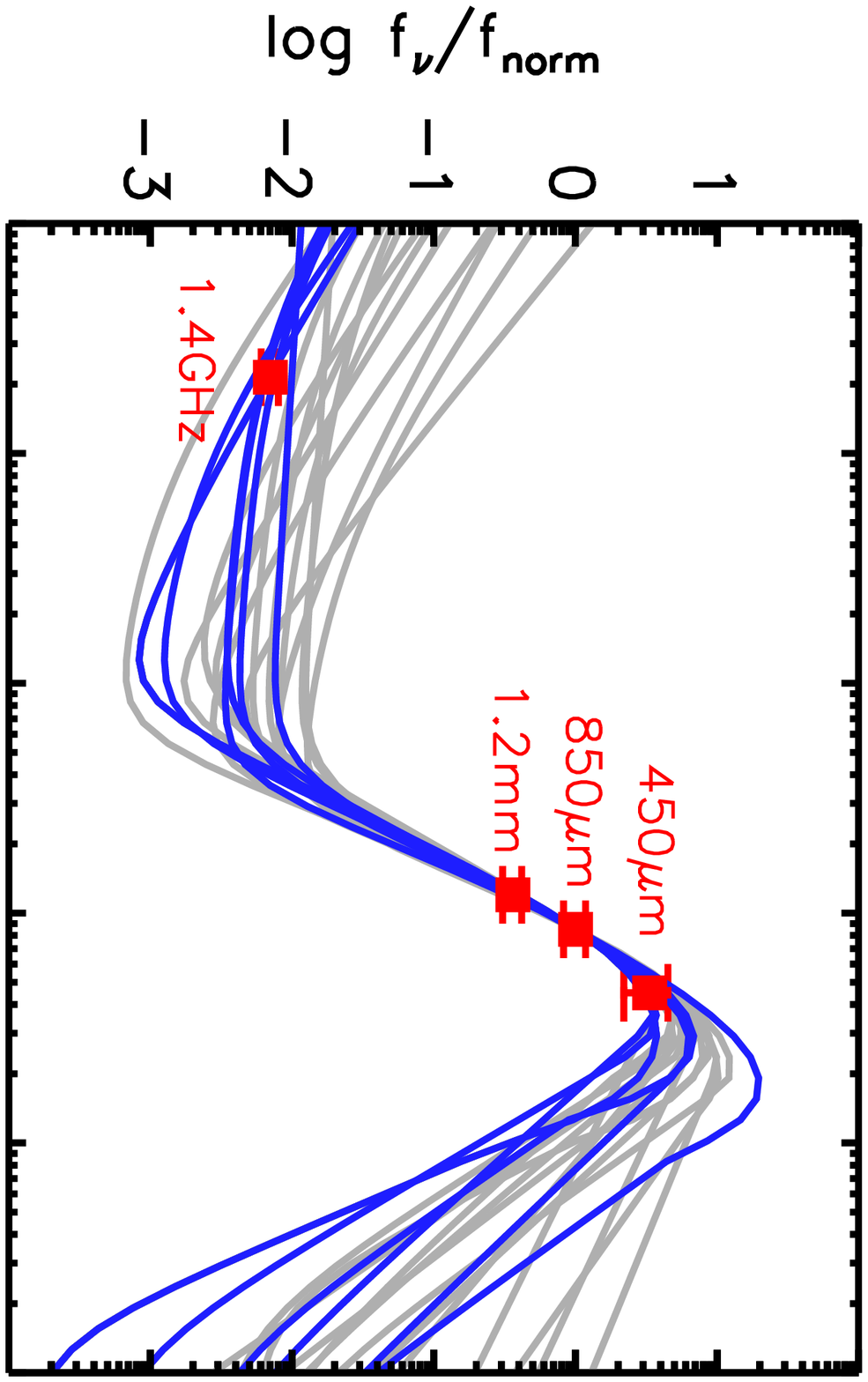}{90} }

    \vspace*{-3.95cm}
    \hbox{ \hspace{3cm} \figl{6cm}{26}{26}{573}{737}{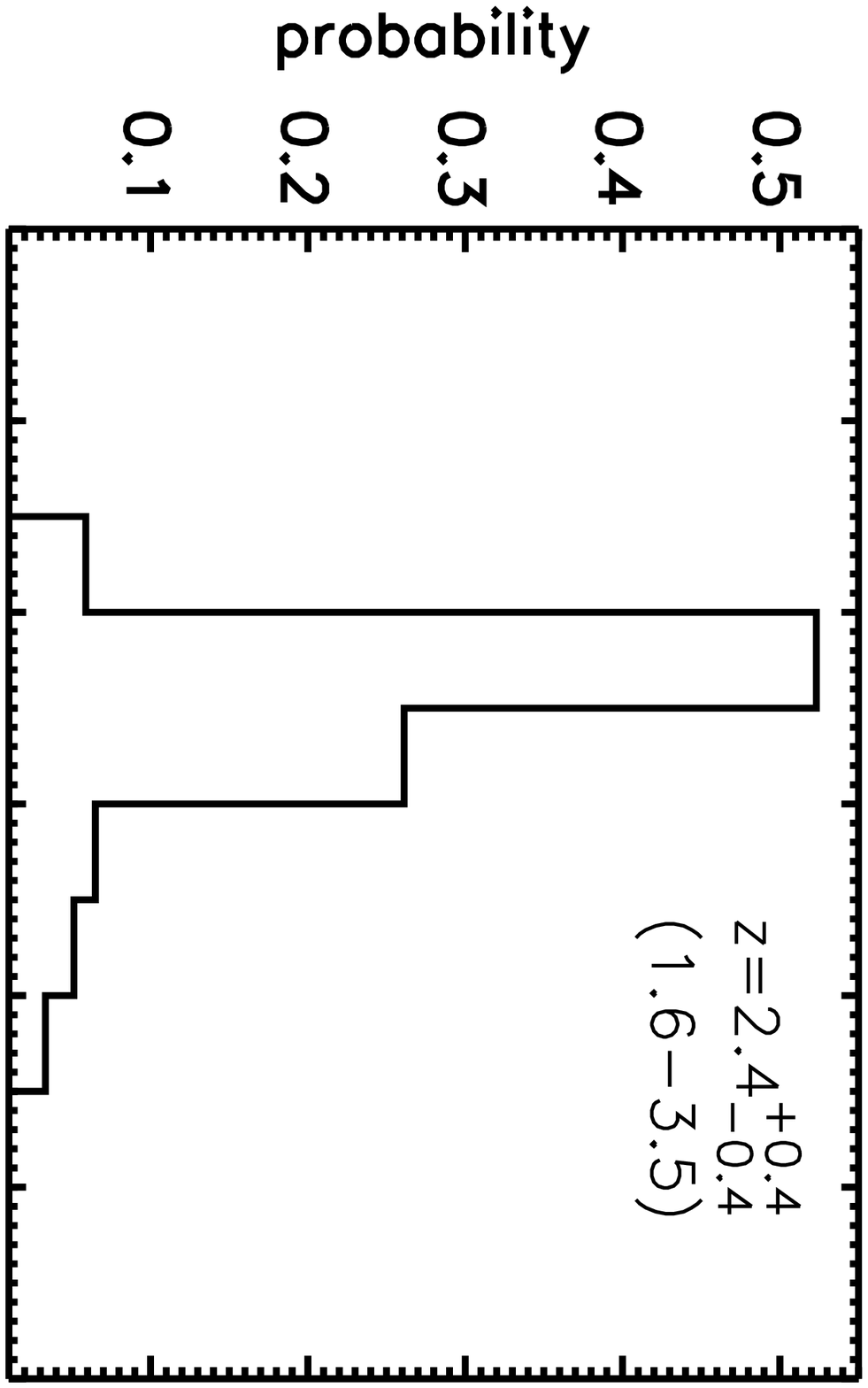}{90} 
           \hspace{-10cm} \figl{6cm}{26}{26}{573}{737}{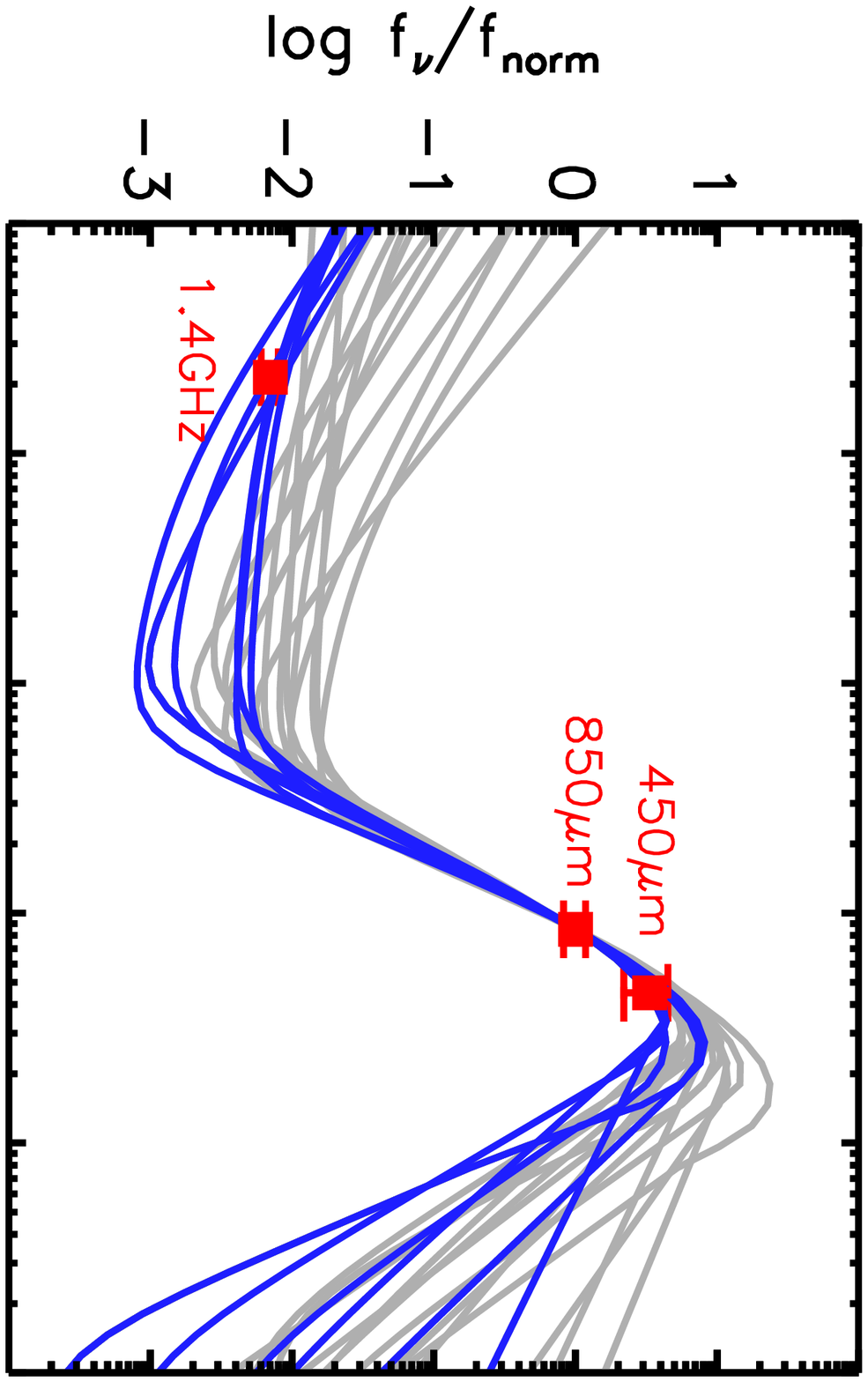}{90} }

    \vspace*{-3.95cm}
    \hbox{ \hspace{3cm} \figl{6cm}{26}{26}{573}{737}{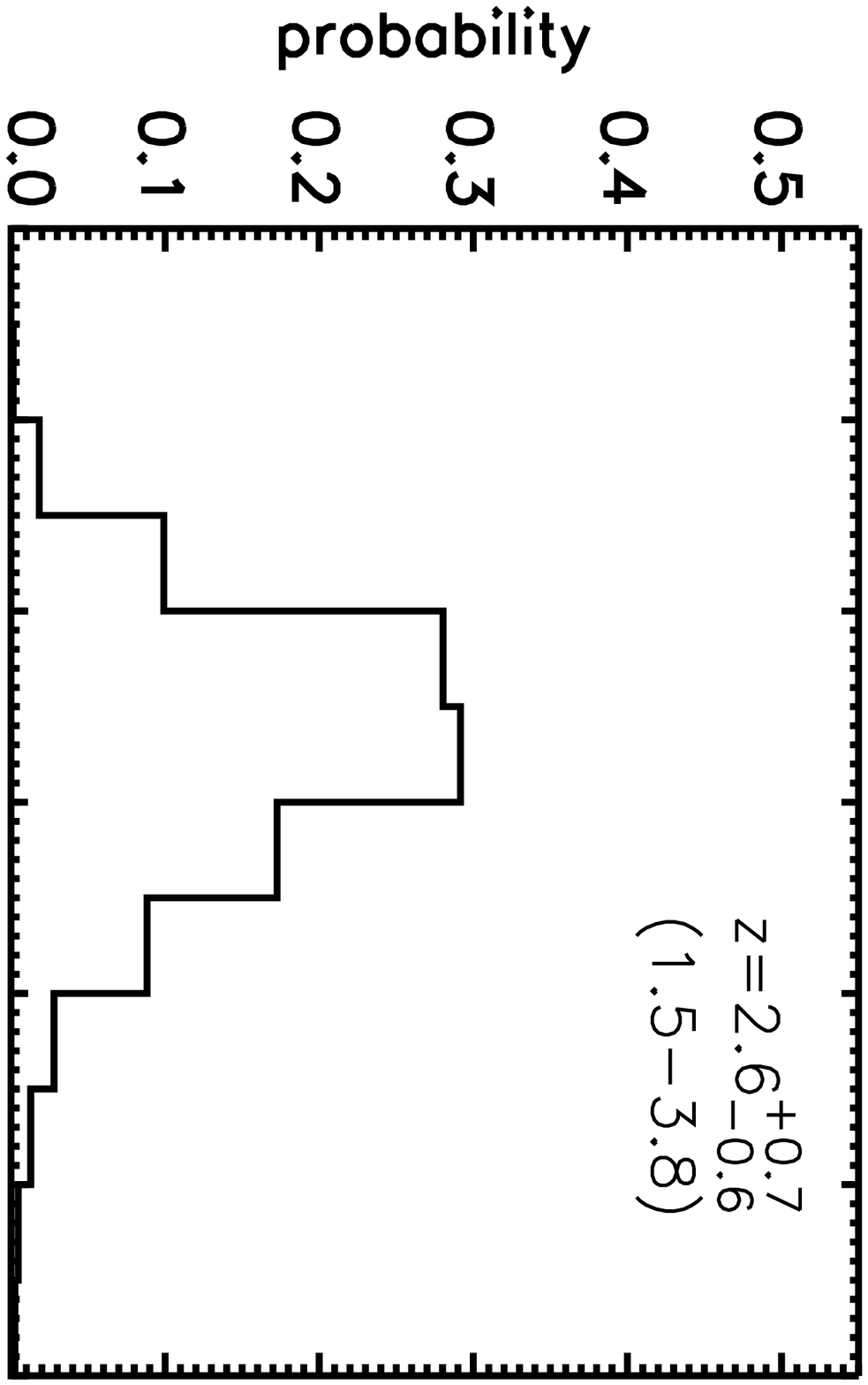}{90} 
           \hspace{-10cm} \figl{6cm}{26}{26}{573}{737}{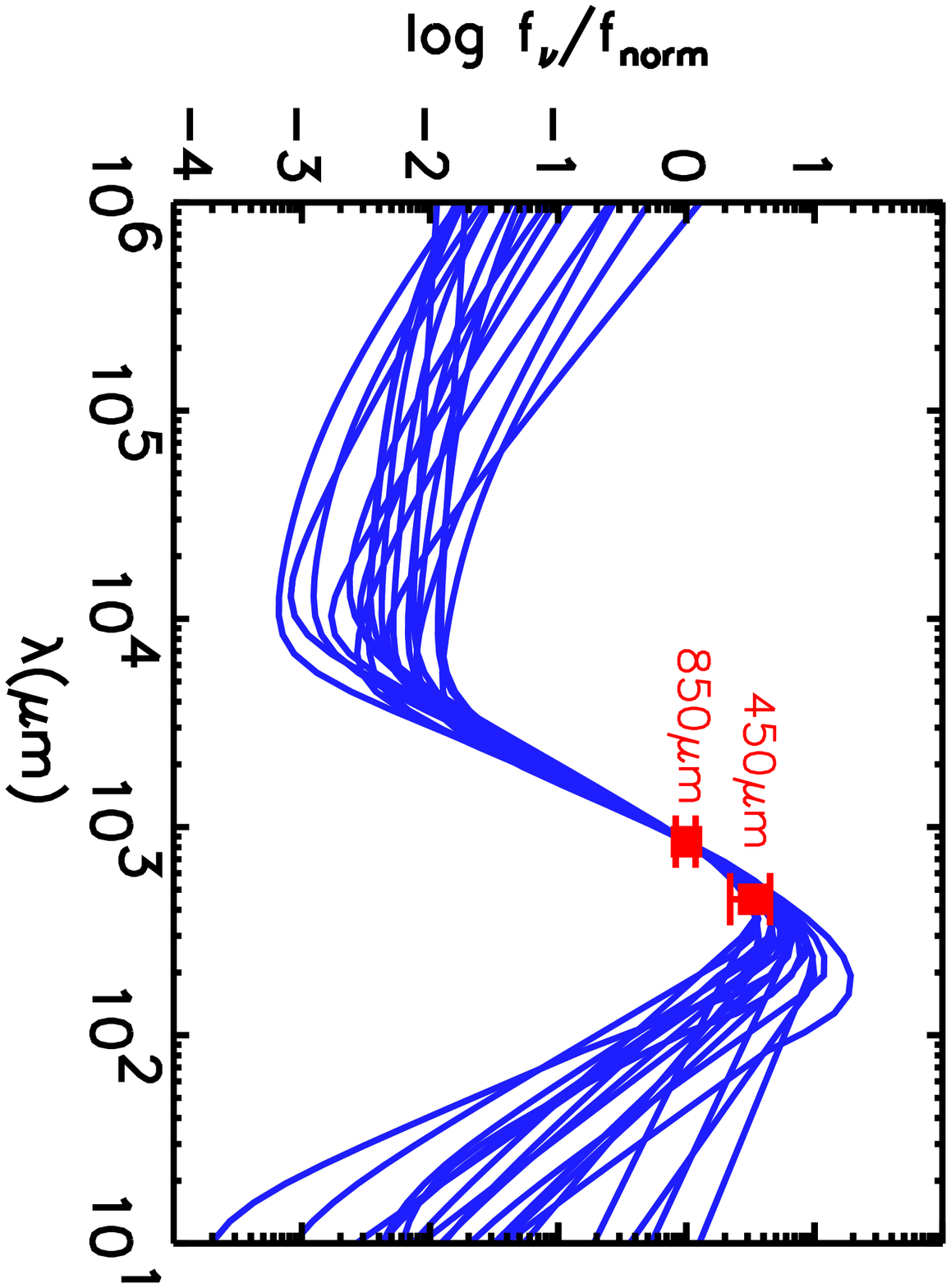}{90} }

    \vspace*{-3.95cm}
    \hbox{ \hspace{3cm} \figl{6cm}{26}{26}{573}{737}{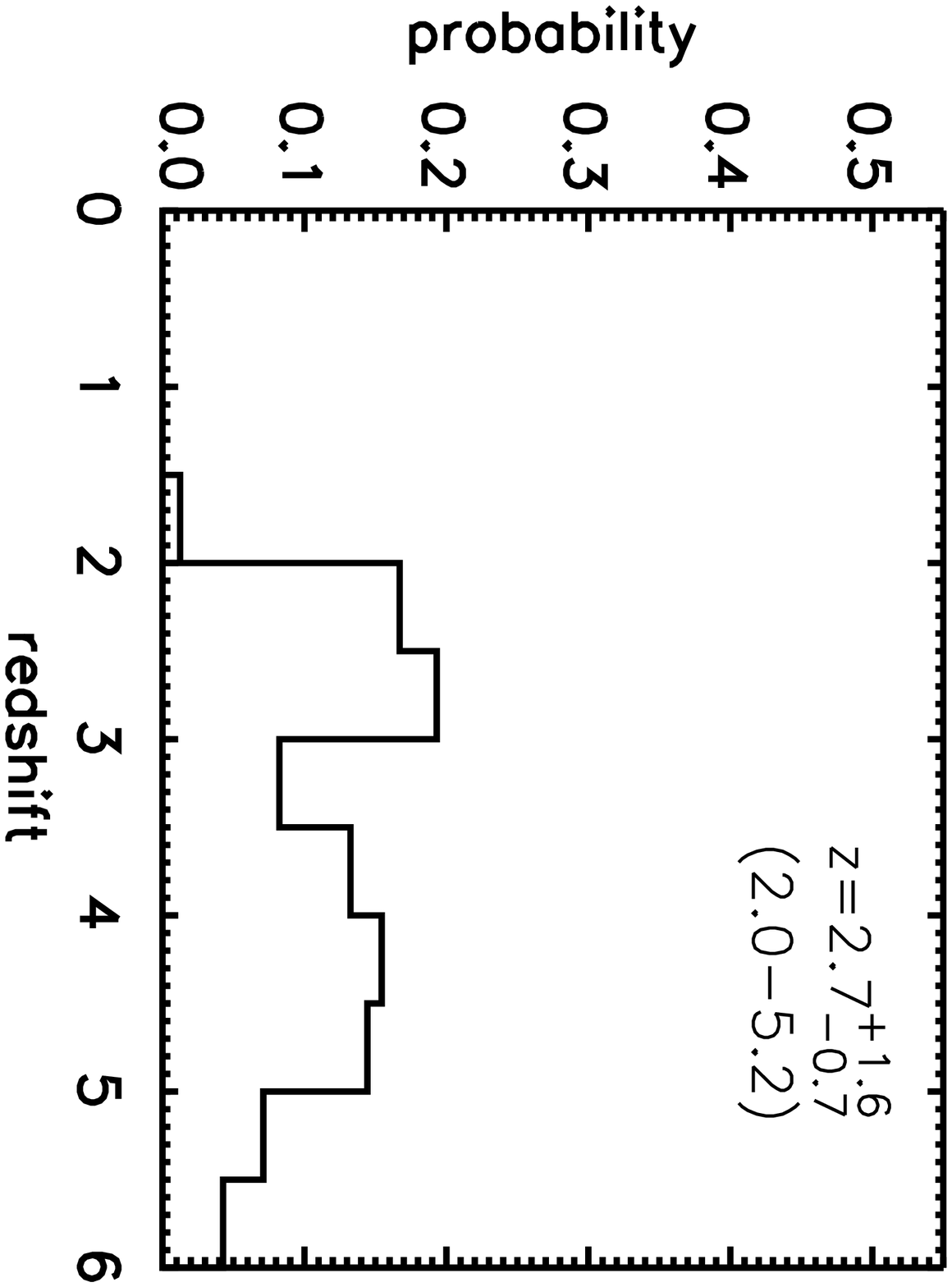}{90} 
           \hspace{-10cm} \figl{6cm}{26}{26}{573}{737}{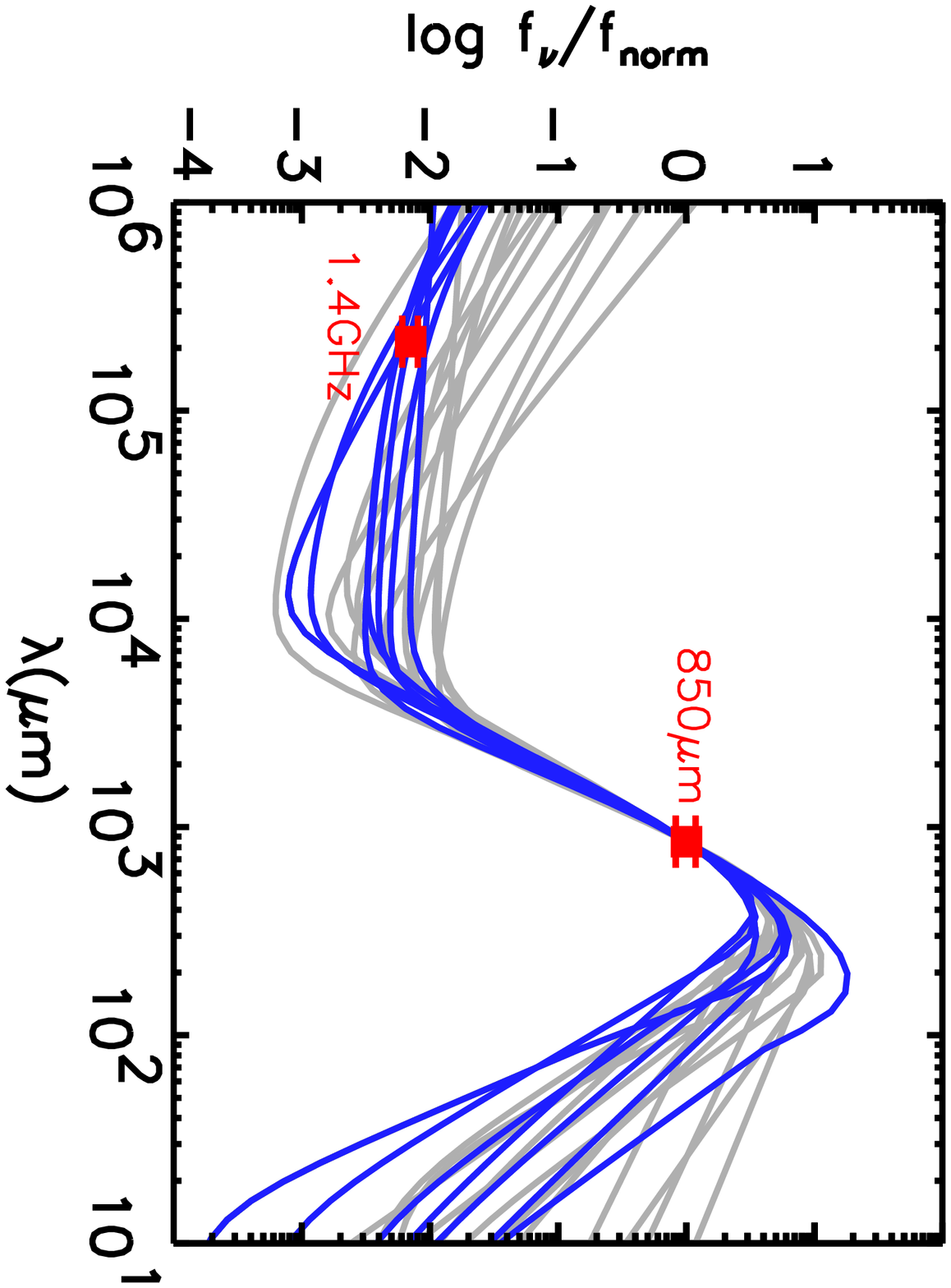}{90} }

\caption{Redshift distributions and compatible template SEDs for LH850.1, when
  different photometry bands  are  taken into account. All estimates
  are based on a Monte Carlo produced with the same evolutionary model (le2), 
  The SED models represented in the right-hand panels have
  been redshifted to the mode of distributions represented in the
  left. Symbol and line codes are as in Fig.~\ref{fig:LH_sed}. }
\label{fig:testref}
\end{figure*}

Fig.~\ref{fig:testref} shows the manner in which the redshift
distribution derived for LH850.1 (using model le2) degrades as
different observational constraints are successively taken into
account. When the 450, 850$\mu$m, 1.2mm and 1.4GHz detections are
included in the analysis, the addition of upper limits to the SED
makes no difference to the derived redshift distribution.  It is also
apparent that the distribution does not change significantly whilst
the detection at 450$\mu$m is used in the analysis, in combination
with at least one additional band in the mm--sub-mm and radio regime.
However, as soon as the 450$\mu$m data is excluded, a high redshift
tail appears.  This is due to the large intrinsic scatter in the
1.4GHz/850$\mu$m ratio amongst the template SEDs which, in the absence
of the 450$\mu$m data, allows galaxies over a wide range of redshifts
to satisfy the colour constraints.  The elimination of the radio data
from the analysis allows more low-redshift sources to be possible
counterparts of the detected source, and hence this flattens the
resulting redshift distribution.

The improvement in constraint on the photometric redshift is
extremely important if it is to be used to determine the tuning of any
broad-band mm--cm spectrograph, in an attempt to detect redshifted
molecular CO-lines. This is discussed further in section\,5.
Higher frequency radio
observations at $\sim 5-15$\,GHz and deep 450$\mu$m observations,
including sensitive upper-limits, are essential
ground-based data that can provide the additional
diagnostic power to constrain redshifts ({\it e.g.} LH850.1).

\subsection{Individual redshift distributions}
Before we calculate the cumulative redshift-distribution for 
the 850$\mu$m sources detected 
in the wide-area ($> 200$~sq. arcmin) blank-field SCUBA surveys, which
include the Lockman Hole and ELAIS N2 regions \cite{scott01,fox01},
and the CUDSS fields \cite{lilly99,eales00,webb02}, it is instructive to
discuss some of the typical redshift distributions that can be derived
from data of different quality (i.e. sensitivity and wavelength
coverage) for individual sources.

Therefore, we compare the redshift
distributions of wide-area  blank-field sub-mm sources with 
those detected in various lensing-cluster
surveys \cite{smail97,smail99,bertoldi00,chapman02}. We also consider
radio, FIR and UV-selected objects with follow-up sub-mm data
\cite{barger00,dscott01,chapman00}, and two bright sub-mm sources with
extensive multi-wavelength observations, HDF850.1
\cite{hughes98,downes99,dunlop02}, and the extremely-red object HR10 \cite{dey99}. 

Tables\,1, 2, and 3 summarize the photometric redshifts of all the
above sources: the modes of their redshift distributions, 68\% and
90\% confidence intervals derived from the six models studied in this 
paper,  and  other 
relevant information. 
Appendix\,A contains the $C-C-z$, $C-f-z$, $P-z$ diagrams and
SEDs of the galaxies listed in Tables 1 and 2, and also a selection of
those listed in Table~3 for model le2, which is one of the less
constraining models considered and serves as an illustration of the
points made here.

When we have the most complete multi-wavelength data-set (rest-frame
radio--sub-mm--FIR, including upper limits) then photometric redshifts
can be constrained to a 68\% confidence band of width $\Delta z \sim
\pm 0.5$ (Table~1 and Fig.A1). Those sub-mm galaxies for which just
two detections, and perhaps some shallow upper-limits are available,
do much worse, with accuracies at a 68\% confidence level of width
$\Delta z \gsim \pm 1$ (Table~2 and Figs.~A2).  A merit of the Monte
Carlo photometric-redshift method is that it can determine redshift
probability distributions even in the case that the galaxies have just
been detected in one band (Table~3 and Fig.~A3), the result being very
flat redshift distributions, that are highly dependent on the assumed
evolutionary model.  This is especially important because half of the
sub-mm sources in the blank-field surveys belong to this category,
although it is worth noting that several of these previously
reported sub-mm sources have now been found to be spurious (Ivison et
al. 2002).

Often, the well constrained distributions are asymmetric, skewed towards lower
redshifts with a high-redshift tail ({\it e.g.} LH850.1, N2850.2, HR10, ... in
Figs.~2 and A1). This is a natural consequence of the ability of the data
(particularly with a deep 450$\mu$m detection or upper limit, $< 30$\,mJy) to
strongly reject the possibility of sub-mm galaxies lying at $z \lsim 2$,
whilst the same observational data cannot reject the possibility that the
sub-mm galaxies lie at a much higher redshift, $z \gg 3$. Examples of the power
with which the 450$\mu$m detections or upper-limits can reject $z\lsim2$ can
be seen in the SEDs of LH850.1 and HDF850.1 (Figs.~3 and A1), 
respectively.  On the
other hand, the bright ISOPHOT detections at 170$\mu$m ($> 120$mJy) make a
strong case that the sources lie at $z\lsim 2$, for example N1--40,
N1--64 (Fig.~A1)
and N1--8 (Fig.~A2), whilst the ISOPHOT non-detections are too shallow to
offer any additional constraints on the SEDs of the majority of the sub-mm
population (e.g.  LH850.1, HDF850.1 in Figs.~3 and A1).

It is also possible to obtain a bimodal redshift probability
distribution for some galaxies (e.g. SMMJ00266+1708 in Fig.~A1, BCR33
in Fig.~A2), an effect created by two or more template SEDs
reproducing the same observed colours at very different redshifts.

The figures in appendix\,A show the insensitivity of the
850$\mu$m/1.4GHz ratio to redshift when the sources are at $z\gsim 2$
(a point also made by CY99, CY00). The redshift distributions in these
cases are usually shallow, with long high-redshift tails (e.g. BCR11,
BCR13, MMJ154127+6615, in Fig.~A2), although exceptions also occur
(SMMJ16403+4644 in Fig.~2A).  The distributions are even broader in
the cases when only a sub-mm detection and a radio upper-limit are
available (Fig.~A3).  All these redshift distributions are
narrower for those evolutionary models with a strong decline beyond
$z\sim 2-3$, as can be seen from the measurements in Tables~2 and 3,
especially for lde1 which essentially contains no sources at
high-$z$.

A quick browse through Appendix~A illustrates the variety of template
SEDs that reproduce the colours of an individual sub-mm galaxy, which
lends support to the use of a template library in the derivation of
the photometric redshifts. A few sub-mm sources, however, still cannot
be reproduced at any redshift with the template SEDs available
(e.g. LH850.12, SMMJ16403+46437).
A possible explanation for this problem is the
contribution of AGN emission to the radio-fluxes at a level not
included in the SEDs of the template library, or a mis-identification
of a sub-mm galaxy with a nearby radio-source (see also section\,4).
A wider range of SEDs, including modest fractions of galaxies with an excess 
of radio-emission, 
could partially solve  this problem but, on  the other hand, it
would also introduce an extra error term in the redshift computation for the
rest of  the sub-mm galaxies.
Other sub-mm galaxies have only a small
representation of analogs among the mock galaxies (e.g. N1-40,
SMMJ14009+0252),  most probably because they are rare
  objects in the Universe, either by their intrinsically high luminosity or 
their low redshift.
This latter point is illustrated by the
rarity of mock galaxies in the corresponding $C-C-z$ or $C-f-z$
diagrams, assuming any of the evolutionary scenarios considered in
this paper.

\subsubsection{Accuracy of the photometric-redshift method}

Fig.~8 shows the comparison of photometric and
published spectroscopic redshifts for 8 galaxies: 
in order of increasing \hbox{spectroscopic} redshift,
N1--40, CUDSS14.18, N1--64, 
SMMJ02399--0134, HR10, SMMJ14011+0252, SMMJ02399$-$0136, and W$-$MM11.

The two galaxies which depart most from the $z_{\rm phot}=z_{\rm
spec}$ line are SMMJ02399-0134 and W$-$MM11.  The first has
a photometric redshift calculated only on the basis of its
1.4\,GHz/850$\mu$m colour (Fig.~A2), and the latter on a 850$\mu$m
detection with a $450\mu$m upper limit (Fig.~A3). These two objects
illustrate the large confidence intervals that have to be considered
when the redshifts are derived from a single colour. Specifically,
SMMJ02399$-$0134 illustrates the uncertainties attached to the large
dispersion of 1.4GHz/850$\mu$m colours implied from a local sample of
galaxies.  The agreement between spectroscopic and photometric
redshifts for the remaining six galaxies (with more than one colour
available) is remarkably good. The asymmetry of 
the error bars is a reflection of the
skewness of their respective
$P-z$ distributions (Figs. A1, A2). These sources
illustrate the importance of obtaining the description of the whole redshift
distribution instead of relying on measurements of the median, which
departs significantly from the most probable redshift in skewed distributions.

A larger
sample of spectroscopic redshifts is obviously 
essential to assess the statistical
significance of the goodness of this technique,
and the accuracy of the error bars we derive. For the sources
with more than one measured colour, these errors are mainly the 
result of the scatter in the shapes of the template SEDs. A larger
sample of spectroscopic redshifts would allow us to fine-tune the
selection  of SEDs and priors in order to trim the error bars if, for
instance, the modes of the distributions do not populate the envelope of
error bars implied from the analysis. The six best constrained
redshift distributions (shown in black in Fig. 7) do show a tendency to cluster more
around the $z_{\rm spec}=z_{\rm phot}$ line than the collective
extension of their error bars. However, fine-tuning the inputs of the
Monte-Carlo at this early stage of confirmation of  the photometric 
redshifts is premature.

\begin{figure}
    \vspace*{-2.5cm}
\hspace*{9.5cm} \figl{8cm}{26}{26}{573}{737}{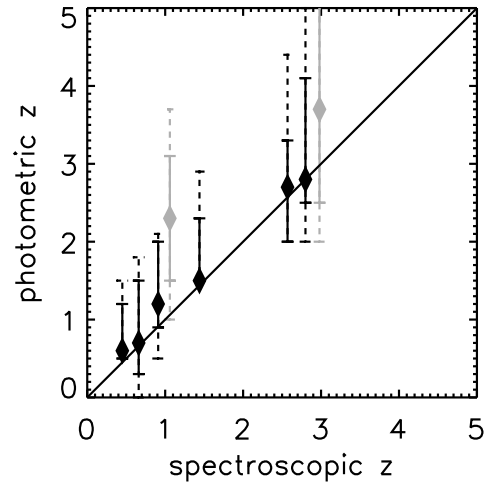}{90}
\caption{Comparison of the photometric-redshift estimates and the true
redshifts of sub-mm sources with published optical/IR spectroscopy.  The
diamonds represent the modes of the redshift distributions for le2. The solid
and dashed error bars represent 68\% and 90\% confidence intervals,
respectively. Sources represented in black (in increasing redshift:
N1--40, CUDSS14.18, N1--64, HR10, SMMJ14011+0252, and SMMJ02399$-$0136) have
photometric redshifts derived from a combination of colours, and are
the most precise. Sources represented in grey (in increasing redshift:
SMMJ02399$-$0134, and W$-$MM11) have photometric redshifts derived
from only one colour (or limit), and have shallow probability
distributions, still compatible with their spectroscopic redshifts.
Note that the asymmetry of the error-bars is a reflection of the
asymmetry of the redshift probability distributions, which in cases
(CUDSS14.18 and HR10) are significantly
 skewed.  }
\label{photz_z}
\end{figure}

\begin{table*}
 \begin{minipage}{170mm}
\begin{center}
\caption{Sources detected in three or more passbands at a
$\geq 3\sigma$ level. Column 1 gives
the name of the source; column 2 the
references for the SED data; columns 3-8 
the mode of the photometric-redshift distributions and 68\% 
confidence intervals for different evolutionary models, and in
parenthesis below, the corresponding 90\% confidence intervals; and column 9
 notes for each object.
The references are
coded as follows: 
                      (1) Lutz et al. 2001,
                      (2) Scott et al. 2002, 
                      (3) Ivison et al. 2002, 
                      (4) Downes et al. 2001, 
                      (5) Dunlop et al. 2002;
                      (6) Eales et al. 2000, 
                      (7) Smail et al. 1999, 
                      (8) Frayer et al. 2000, 
                      (9) Ivison et al. 1998, 
                      (10) Ivison et al. 2000, 
                      (11) Dey et al. 1999,
                      (12) Chapman et al. 2002b.
                      (13) Cowie et al. 2002
The $A$ in the notes denotes the
amplification factor reported in the literature for the lensed
sources.} 
\label{3filters}
\begin{tabular}{llccccccl}
\hline
%  object & ref & $z_{\rm phot}^{\rm le1}$ & 68\% CL & 90\% CL & notes \\
  object & ref & $z_{\rm phot}^{\rm le1}$ & $z_{\rm phot}^{\rm
    le1L13}$ & $z_{\rm
    phot}^{\rm le2}$ & $z_{\rm phot}^{\rm le2L13}$ & $z_{\rm
    phot}^{\rm lde1}$ &
    $z_{\rm phot}^{\rm lde2}$ & notes \\
\hline
LH850.1          & 1,2,3 & 2.6$\pm^{0.4}_{0.6}$ & 
                           2.6$\pm^{0.4}_{0.5}$ & 
                           2.6$\pm^{0.4}_{0.5}$ & 
                           2.6$\pm^{0.4}_{0.5}$ & 
                           2.7$\pm0.3$ & 
                           2.7$\pm^{0.3}_{0.2}$ & \\
                 &       & (2.0--4.1) & (2.0--3.8) & (2.0--3.7) & 
			   (2.0--3.9) & (2.0--3.1) &  (2.3--3.5) & \\
LH850.3          & 2,3   & 2.7$\pm0.2$ &
                           2.7$\pm0.2$ & 
                           3.0$\pm^{0.0}_{0.5}$ & 
                           2.5$\pm^{0.0}_{1.3}$ & 
			   2.9$\pm^{0.6}_{1.2}$ & 
			   3.0$\pm^{0.0}_{0.5}$ &  \\
                 &       & (2.1--3.0) & (1.8--3.0) & (2.1--3.0) & 
                           (1.0--3.0) & (1.0--3.5) &  (2.1--3.0) & \\
LH850.8          & 2,3   & 3.8$\pm^{1.2}_{1.4}$ &  
                           3.8$\pm^{1.3}_{1.2}$ & 
                           3.7$\pm^{1.5}_{0.7}$ & 
			   3.8$\pm^{1.2}_{0.8}$ & 
			   3.5$\pm^{0.0}_{0.6}$ & 
			   3.0$\pm^{0.0}_{0.9}$ & \\
                 &       & (2.0--6.5) & (1.6--6.0) & (2.3--6.0) & 
                           (2.0--5.3) & (2.6--3.5) &  (2.0--4.3) & \\
LH850.12         & 2,3   & 0.6$\pm^{0.3}_{0.1}$ &  
                           0.6$\pm^{0.3}_{0.1}$ & 
                           0.2$\pm^{0.8}_{0.2}$ & 
			   0.2$\pm^{0.6}_{0.2}$ & 
			   0.7$\pm0.2$ & 
			   1.0$\pm0.5$ & 
    radio-loud QSO$^{(3)}$, $z_{\rm phot}$ is\\
                 &       & (0.1--1.0) & (0.1--1.0) & (0.0--1.3) & 
                           (0.0--1.2) & (0.5--1.0) & (0.8--1.5) & 
   unreliable \\
N2850.1          & 2,3   & 2.9$\pm0.6$ &  
                           2.8$\pm^{0.7}_{0.4}$            & 
                           2.8$\pm^{0.7}_{0.5}$ & 
			   2.8$\pm^{0.3}_{0.8}$ & 
			   2.9$\pm^{0.3}_{0.4}$ & 
			   2.7$\pm^{0.3}_{0.4}$ &   \\ 
                 &       & (2.0--4.1) & (2.0--3.7) & (2.0--4.1) & 
                           (2.0--5.3) & (2.4--3.5) &    (2.0--3.5) &  \\
N2850.2          & 2,3   & 2.3$\pm^{0.6}_{0.3}$ &
                           2.3$\pm^{0.5}_{0.3}$ & 
                           2.3$\pm^{0.6}_{0.3}$ & 
			   2.3$\pm^{0.5}_{0.3}$ & 
			   2.3$\pm0.3$ & 
			   2.6$\pm^{0.4}_{0.1}$ &   \\ 
                 &       & (1.5--3.4) & (1.5--3.2) & (1.5--3.5) & 
                           (1.5--3.0) & (2.0--3.4) &    (2.1--3.0) & \\
N2850.4          & 2,3   & 2.8$\pm^{1.7}_{0.3}$ & 
                           2.7$\pm0.7$ & 
                           2.9$\pm^{1.2}_{0.9}$ & 
                           2.8$\pm^{0.9}_{0.8}$ & 
			   2.8$\pm^{0.5}_{0.3}$ & 
			   2.7$\pm^{0.3}_{0.7}$ &   \\ 
                 &       & (1.3--4.5) & (2.0--4.5) & (1.2--4.5) & 
                           (1.6--4.5) & (2.2--3.5) & (1.5--3.4) & \\
HDF850.1         & 4,5   & 4.2$\pm^{0.5}_{0.7}$ &
                           4.1$\pm0.6$ & 
                           4.1$\pm0.6$ & 
			   4.0$\pm^{0.7}_{0.5}$ & 
			   3.5$\pm^{0.0}_{0.6}$ & 
			   4.0$\pm^{0.5}_{0.6}$ &  
$A=3$ \\ 
                 &       & (3.3--5.5) & (3.1--5.0) & (3.0--5.1) & 
			   (3.0--5.0) & (2.7--3.5) &  (3.2--5.0) &\\
CUDSS14.1        & 6     & 3.8$\pm^{0.2}_{0.7}$ & 
                           3.8$\pm^{0.8}_{0.7}$ & 
                           3.8$\pm^{0.7}_{0.8}$  & 
			   3.7$\pm0.8$ & 
			   2.9$\pm{0.4}$ & 
			   2.8$\pm^{1.1}_{0.8}$ &   \\ 
                 &       & (2.5--5.2) & (2.5--5.5) & (2.0--4.9) & 
                           (2.2--5.0) & (2.1--3.5) &  (2.0--4.8) & \\
SMMJ00266+1708   & 7,8   & 2.7$\pm^{2.2}_{0.2}$ & 
                           2.7$\pm^{2.0}_{0.2}$ & 
                           2.7$\pm^{1.6}_{0.7}$ & 
			   2.8$\pm^{2.2}_{0.2}$ & 
			   2.8$\pm^{0.5}_{0.3}$ & 
			   2.9$\pm^{0.5}_{0.4}$ &  $A=1.6$  \\ 
                 &       & (2.1--5.5) & (2.2--5.5) & (2.0--5.4) & 
                           (1.0--5.3) & (2.2--3.5) &  (2.5--5.3) &\\
SMMJ02399$-$0136 & 8,9   & 3.5$\pm^{0.0}_{1.0}$ &
                           3.1$\pm^{0.8}_{0.6}$ & 
                           3.9$\pm 0.9$ & 
			   3.1$\pm^{0.7}_{0.6}$ &  
			   3.5$\pm^{0.0}_{0.8}$ &
			   3.8$\pm^{1.7}_{0.8}$ &
                           $z_{\rm spec}=2.8$, $A=2.4$, \\  
                 &       & (2.3--5.0) & (2.0--4.5) & (2.5--5.0) & 
			   (2.5--4.6) & (2.2--3.5) &  (2.4--5.0) &
    AGN, 3.4cm too low? \\                                         
                 &       & 2.8$\pm0.7$ &  
                           2.9$\pm0.6$          & 
                           2.8$\pm^{1.3}_{0.3}$ & 
			   2.8$\pm^{0.7}_{0.5}$ & 
			   2.8$\pm^{0.4}_{0.3}$ & 
			   2.8$\pm^{0.8}_{0.3}$ &
    disregarding the 3.4cm upper limit\\  
                 &       & (2.0--4.5) & (2.0--4.4) & (2.0--5.0) & 
                           (2.0--4.3) & (2.0--3.5) & (2.0--4.4) & \\                                         
SMMJ09429$+$4658 & 7,13    & 2.8$\pm^{1.7}_{0.3}$ &  
                           2.8$\pm0.6$ & 
                           2.8$\pm^{1.5}_{0.3}$ & 
			   2.8$\pm^{0.7}_{0.4}$ & 
			   3.0$\pm^{0.5}_{0.1}$ & 
			   2.7$\pm^{0.3}_{0.6}$ & $A=1.5$\\
                 &       & (2.0--5.6) & (1.9--4.0) & (2.0--4.9) & 
                           (1.9--4.0) & (2.3--3.5) & (2.0--4.3) & \\                                         
SMMJ14009+0252   & 10    & 4.4$\pm^{0.7}_{0.9}$ &
                           4.1$\pm^{0.8}_{0.6}$  & 
                           4.1$\pm0.8$  & 
			   4.1$\pm{0.8}$ & 
			   3.5$\pm^{0.0}_{0.9}$ & 
			   4.1$\pm^{0.4}_{1.2}$  &  $A=1.5$\\
                 &       & (2.7--5.5) & (3.0--5.1) & (3.1--5.0) & 
                           (2.5--5.2) & (2.1--3.5) &    (2.0--5.0) &
    AGN contribution? \\                                         
SMMJ14011+0252   & 8     & 2.8$\pm^{0.5}_{0.8}$ &  
                           2.8$\pm^{0.5}_{0.8}$ & 
                           2.7$\pm^{0.6}_{0.7}$ & 
			   2.7$\pm^{0.6}_{0.7}$ & 
			   2.7$\pm^{0.6}_{0.2}$ & 
			   2.7$\pm^{0.3}_{0.4}$& 
                           $z_{\rm spec}=2.57$,    $A=3.$\\
                 &       & (2.0--4.5) & (2.0--4.5) & (2.0--4.4) & 
                           (2.0--4.5) & (2.3--3.5) &    (2.0--3.5) & \\                                         
HR10             & 11     & 1.5$\pm^{1.0}_{0.0}$ &  
                            1.5$\pm^{0.5}_{0.0}$ & 
                            1.5$\pm^{0.8}_{0.0}$ & 
			    1.5$\pm^{0.7}_{0.0}$ & 
			    1.5$\pm^{1.0}_{0.0}$ & 
			    1.5$\pm^{0.7}_{0.0}$ & 
                            $z_{\rm spec}=1.44$ \\
                 &       & (1.5--3.1) & (1.5--2.5) & (1.5--2.9) & 
			   (1.5--2.6) & (1.5--3.1) &   (1.5--2.4) & \\                                         
N1-40            & 12     & $1.2\pm0.2$ &  
                            0.7$\pm^{0.6}_{0.1}$  & 
                            0.6$\pm^{0.6}_{0.1}$ & 
			    0.6$\pm0.2$ & 
			    0.7$\pm0.2$ & 
			    1.0$\pm^{0.5}_{0.0}$ & 
			    $z_{\rm spec}=0.45$ \\
                 &       & (0.6--1.5) & (0.5--1.5) & (0.5--1.5) & 
                           (0.3--1.0) & (0.5--1.0) &    (0.5--1.5) & \\                                         
N1-64            & 12    & 1.2$\pm^{0.3}_{0.8}$ & 
                           1.2$\pm0.3$ & 
                           1.2$\pm^{0.8}_{0.3}$ & 
			   1.2$\pm^{0.3}_{0.6}$ & 
			   1.2$\pm^{0.3}_{0.6}$ & 
			   1.2$\pm^{0.6}_{0.2}$ & 
                           $z_{\rm                 spec}=0.91$ \\
                 &       & (0.5--1.8) & (0.5--1.5) & (0.5--2.1) & 
			   (0.5--1.8) & (--) &    (0.7--2.0) & \\

\hline
\end{tabular}
\end{center}
\end{minipage}
\end{table*}

\begin{table*}
 \begin{minipage}{140mm}
\begin{center}
\caption{Sources detected in two pass-bands at $\geq 3\sigma$
level. The columns are
identical to those described in Table\,1.
The reference
codes are as follows: 
        (1) Scott et al. 2002,
        (2) Ivison et al. 2002
        (3) Barger et al. 2000,                 
        (4) Eales et al. 2000, 
        (5) Webb et al. 2002,
        (6) Smail et al. 1999,
        (7) Chapman et al. 2002a,
        (8) Bertoldi et al. 2000,
        (9) Dannerbauer et al. 2002,
        (10) Chapman et al. 2002b.
} \label{2filters}
\vspace*{-0.4cm}
\begin{tabular}{llccccccl}
\hline
%  object & ref & $z_{\rm phot}^{\rm le1}$ & 68\% CL & 90\% CL & notes \\
  object & ref & $z_{\rm phot}^{\rm le1}$ & $z_{\rm phot}^{\rm
    le1L13}$ & $z_{\rm
    phot}^{\rm le2}$ & $z_{\rm phot}^{\rm le2L13}$ & $z_{\rm
    phot}^{\rm lde1}$ &
    $z_{\rm phot}^{\rm lde2}$ & notes \\
\hline
LH850.6         & 1,2   & 2.7$\pm^{2.0}_{0.7}$ & 
                          2.7$\pm^{2.0}_{0.7}$ & 
                          2.7$\pm^{1.8}_{0.7}$ & 
                          2.7$\pm^{1.8}_{0.7}$ & 
			  2.7$\pm^{0.6}_{0.2}$ & 
                          2.8$\pm^{0.2}_{0.6}$ & \\
& & (2.0--6.4) & (2.0--6.2) & (2.0--5.4) & 
    (2.0--5.4) & (2.3--3.5) & (2.0--4.4) & \\
LH850.7         & 1,2     & 3.6$\pm^{1.9}_{0.8}$ &  
                          3.5$\pm^{0.5}_{1.8}$ & 
                          3.6$\pm^{0.4}_{1.9}$ & 
                          3.5$\pm^{0.5}_{1.9}$ & 
			  3.5$\pm^{0.0}_{1.0}$ & 
			  2.5$\pm^{0.2}_{0.5}$ & \\
& & (1.6--5.5) & (1.5--5.1) & (1.5--5.3) & 
    (1.5--5.3) & (1.7--3.5) & (1.5--3.8) & \\
LH850.14        & 1,2     & 2.5$\pm^{1.9}_{0.4}$ & 
                            2.4$\pm^{2.0}_{0.4}$ & 
			    2.4$\pm^{1.9}_{0.4}$ & 
			    2.3$\pm^{2.0}_{0.3}$ & 
			    2.7$\pm^{0.3}_{0.5}$ & 
			    3.0$\pm^{0.0}_{0.7}$ & \\
  & & (1.5--6.3) & (1.5--5.5) & (2.0--5.4) & 
      (2.0--5.4) & (2.0--3.4) & (2.0--3.5)  & \\
LH850.16        & 1,2     & 3.6$\pm^{1.0}_{1.6}$ & 
                            3.5$\pm^{1.0}_{1.5}$ &
			    3.3$\pm^{1.1}_{1.3}$ & 
			    3.6$\pm^{0.8}_{1.5}$ &
			    2.8$\pm^{0.5}_{0.3}$ & 
			    3.5$\pm^{0.0}_{0.8}$& \\
  & & (1.5--6.0) & (1.5--6.1) & (2.0--5.5) &
      (2.0--5.4) & (2.0--3.5) & (1.5--3.5) & \\
LH850.18        & 1,2     &  3.2$\pm{1.1}$ & 
                             3.4$\pm^{1.0}_{1.4}$ & 
			     3.0$\pm^{1.1}_{1.0}$ & 
			     3.2$\pm^{1.0}_{1.2}$ & 
			     3.2$\pm^{0.3}_{0.7}$ & 
			     3.0$\pm^{0.0}_{0.7}$ & \\
  & & (1.5--5.8) & (1.5--5.6) & (1.5--4.9) &
      (1.8--4.9) & (2.0--3.5) & (2.0--3.3)& \\
N2850.7         & 1,2     &  2.8$\pm^{1.5}_{0.8}$ & 
                             2.8$\pm^{1.2}_{0.6}$ & 
			     3.0$\pm1.0$ & 
			     3.0$\pm0.9$ & 
			     3.0$\pm^{0.3}_{0.5}$ & 
			     2.5$\pm0.5$ & \\
  & & (1.9--5.5) & (2.0--5.3) & (1.5--4.9) &
      (1.7--5.0) & (2.5--3.5) & (2.0--4.3) & \\
N2850.13        & 1,2     &  2.7$\pm^{0.8}_{1.0}$ &
                             3.0$\pm^{0.5}_{1.5}$   & 
			     2.7$\pm^{0.8}_{1.0}$ & 
			     2.8$\pm^{0.7}_{1.2}$ & 
			     2.8$\pm^{0.5}_{0.3}$ & 
			     2.5$\pm^{0.3}_{0.5}$ & \\
  & & (1.5--5.0) & (1.5--5.0) & (1.5--4.6) & 
      (1.5--4.8) & (2.0--3.5) & (2.0--3.5) & \\
BCR7            & 3       &  2.7$\pm^{0.8}_{1.1}$ & 
                             2.7$\pm^{0.8}_{1.0}$ & 
			     2.2$\pm^{1.1}_{0.7}$ & 
			     2.7$\pm^{0.6}_{1.2}$ & 
			     2.8$\pm^{0.5}_{0.3}$ & 
			     2.4$\pm0.4$ & \\ 
  & & (1.5--4.8) & (1.5--4.7) & (1.5--4.8) &
      (1.5--4.7) & (1.9--3.5) & (1.5--3.2) & \\
BCR11           & 3       & 2.6$\pm0.9$  & 
                            2.7$\pm{0.8}$ & 
			    2.7$\pm^{0.8}_{0.9}$ & 
			    2.6$\pm^{0.7}_{1.1}$ & 
			    2.7$\pm^{0.6}_{0.2}$ &
			    2.4$\pm0.4$ & \\ 
  & & (1.5--4.7) & (1.5--4.7) & (1.5--4.6) &
      (1.5--4.7) & (2.1--3.5) & (1.5--3.0) & \\
BCR13           & 3       &  2.3$\pm^{1.9}_{0.3}$ & 
                             2.3$\pm^{2.0}_{0.3}$  & 
			     2.3$\pm^{1.7}_{0.3}$ & 
			     2.3$\pm^{1.7}_{0.3}$ & 
			     2.8$\pm^{0.7}_{0.3}$ & 
			     2.9$\pm^{0.1}_{0.6}$ & \\
  & & (1.5--5.5) & (1.5--5.6) & (1.5--4.7) &
      (1.5--4.7) & (2.2--3.5) & (2.0--3.3) & \\
BCR33           & 3       &  2.9$\pm1.0$ & 
                             3.0$\pm1.0$ &  
                             3.0$\pm1.0$ & 
                             3.0$\pm1.0$ & 
			     3.0$\pm^{0.5}_{0.3}$ & 
			     3.0$\pm^{0.0}_{0.7}$ & \\ 
  & & (1.5--5.2) & (1.5--5.3) & (1.5--4.6) &
      (1.5--4.5) & (2.0--3.5) & (2.0--3.0) & \\
BCR49           & 3       & 2.3$\pm^{1.1}_{0.5}$ & 
                            2.4$\pm0.5$  & 
                            2.4$\pm^{1.1}_{0.6}$ & 
                            2.5$\pm^{1.0}_{0.5}$ & 
			    2.7$\pm^{0.3}_{0.7}$ & 
			    2.8$\pm^{0.2}_{0.6}$ & \\ 
  & & (1.6--4.5) & (1.5--4.2) & (1.0--4.0) &
      (1.0--3.7) & (1.7--3.5) & (1.6--3.0) & \\
CUDSS14.3       & 4       &  3.1$\pm0.9$ & 
                             3.1$\pm0.9$ & 
                             3.2$\pm^{0.8}_{1.0}$ & 
                             3.1$\pm0.9$ & 
			     3.5$\pm^{0.0}_{0.8}$ & 
			     3.0$\pm^{0.0}_{0.8}$ & \\
  & & (1.7--5.0) & (1.5--4.8) & (2.0--5.0) &
      (1.7--5.0) & (2.2--3.5) & (2.2--4.1) & \\
CUDSS14.9       & 4       &  2.6$\pm^{0.8}_{0.6}$ &
                             2.6$\pm^{0.9}_{0.6}$ & 
			     2.5$\pm^{0.5}_{1.0}$ & 
                             2.6$\pm^{0.4}_{1.1}$ & 
			     2.7$\pm^{0.8}_{0.2}$ &
			     2.8$\pm^{0.2}_{0.6}$ &  \\
  & & (1.5--4.1) & (1.5--4.0) & (1.5--4.5) &
      (1.5--4.0) & (2.0--3.4) & (1.9--3.5) & \\
CUDSS14.18      & 4       & 0.7$\pm^{0.8}_{0.4}$ &  
                            0.7$\pm^{0.8}_{0.4}$ & 
                            0.7$\pm^{0.8}_{0.4}$ & 
			    0.7$\pm^{1.1}_{0.2}$ & 
			    0.7$\pm0.7$ & 
			    1.2$\pm^{0.5}_{0.7}$ & 
$z_{\rm spec}=0.66$\\
  & & (0.0--1.8) & (0.0--1.8) & (0.0--1.8) &
      (0.0--2.0) & (0.0--1.9) & (0.2--2.0) & \\
CUDSS3.1        & 5       & 2.1$\pm^{0.4}_{0.6}$ & 
                            1.8$\pm^{0.7}_{0.3}$ & 
			    1.7$\pm^{0.7}_{0.2}$ & 
                            1.8$\pm^{0.5}_{0.3}$ & 
			    1.9$\pm0.4$ & 
			    1.8$\pm^{0.4}_{0.3}$ &\\
  & & (1.0--2.5) & (0.7--2.5) & (1.5--2.5) &
      (0.9--2.5)  & (1.5--2.6) & (1.2--2.5) & \\
CUDSS3.8        & 5       & 2.1$\pm^{0.9}_{0.2}$ & 
                            2.0$\pm0.5$ & 
			    2.0$\pm^{0.8}_{0.5}$ & 
                            2.0$\pm0.5$ & 
			    2.6$\pm^{0.4}_{0.2}$  & 
                            1.9$\pm0.4$ & 
high 1.4GHz, AGN?\\
  & & (1.4--3.0) & (1.4--3.0) & (1.0--3.0) &
(1.2--3.0) & (2.0--3.0) & (1.2--2.5)& \\
CUDSS3.9        & 5       &  1.0$\pm^{0.2}_{0.5}$ & 
                             1.0$\pm^{0.4}_{0.0}$  & 
			     1.1$\pm^{0.8}_{0.1}$ & 
			     0.7$\pm^{0.3}_{0.5}$ & 
			     1.0$\pm^{0.4}_{0.0}$ & 
			     1.3$\pm^{0.5}_{0.3}$ & \\
  & & (0.5--1.5) & (1.0--2-0) & (0.5--2.0)&
      (0.0--1.4) & (1.0--1.5) & (0.6--2.0) & \\
CUDSS3.10       & 5       & 2.6$\pm0.8$ & 
                            2.6$\pm0.9$ & 
			    2.5$\pm1.0$ & 
			    2.5$\pm0.9$ & 
			    2.7$\pm^{0.8}_{0.2}$ & 
			    2.5$\pm^{0.0}_{0.9}$ & \\
  & & (1.4--4.5) & (1.3--4.5) & (1.5--4.5)& 
      (1.0--4.2)& (1.9--3.5) & (1.0--2.5) &  \\
CUDSS3.15       & 5       & 2.2$\pm{0.7}$ &
                            2.3$\pm{0.7}$ & 
			    2.2$\pm^{0.9}_{0.7}$ & 
                            2.3$\pm{0.7}$ & 
			    2.5$\pm0.5$ & 
			    2.5$\pm^{0.0}_{0.7}$ & \\
  & & (1.0--3.7) &  (1.5--4.0) & (1.5--4.0) &
      (1.0--3.7) & (1.9--3.5) & (1.2--2.5) & \\
SMMJ02399$-$0134  & 6     &  2.2$\pm0.7$ &  
                             2.3$\pm0.8$ & 
                             2.3$\pm0.8$ & 
			     2.3$\pm0.8$ & 
			     2.6$\pm^{0.7}_{0.5}$ & 
			     2.5$\pm^{0.0}_{0.8}$ & 
$A=2.5$, $z_{\rm spec}=1.06$\\
  & & (1.0--3.6) & (1.4--4.0) & (1.0--3.8) &
      (1.0--3.7)  & (1.7--3.5) & (1.2--2.5) & \\
SMMJ10236+0412  & 7       &  2.3$\pm0.8$ &  
                             2.3$\pm0.8$ & 
                             2.3$\pm0.8$ & 
			     2.3$\pm0.8$ & 
			     2.7$\pm^{0.3}_{0.7}$ & 
			     2.8$\pm^{0.2}_{0.5}$ & $A=3.2$ \\
  & & (1.3--4.0) & (1.3--4.0) & (1.5--4.0) &
      (1.3--4.0) & (2.0--3.5)& (1.7--3.0) & \\
SMMJ10237+0410  & 7       & 2.0$\pm^{1.1}_{0.5}$ & 
                            2.0$\pm^{1.2}_{0.5}$ &
                            2.1$\pm^{1.0}_{0.6}$ & 
			    2.0$\pm^{1.1}_{0.5}$ &
			    2.6$\pm0.6$ & 
			    2.4$\pm0.4$ & $A=3.7$ \\
  & & (1.0--3.5) & (1.0--3.7) & (1.0--3.6) &
      (1.0--3.6) & (1.7--3.5) & (1.6--3.0) & \\
SMMJ16403+4644  & 7       &  2.2$\pm0.8$ & 
                             2.2$\pm0.8$ & 
			     2.1$\pm^{0.4}_{1.1}$ & 
                             2.2$\pm0.7$ & 
			     2.4$\pm^{0.6}_{0.8}$& 
                             2.5$\pm0.4$ & 			    
$A=3.0$ \\
  & & (0.5--3.1) & (0.8--3.5) & (1.0--3.5) &
      (0.5--3.5) & (1.2--3.5) & (1.6--3.0) & \\
SMMJ16403+46440 & 7       &  2.8$\pm0.7$ & 
                             2.8$\pm0.7$ & 
                             2.7$\pm0.8$ & 
			     2.8$\pm0.7$ & 
			     2.8$\pm^{0.6}_{0.3}$ & 
			     2.9$\pm^{0.1}_{0.7}$ & 
$A=3.6$ \\
  & & (1.5--4.1) & (1.5--4.1) & (1.5--4.1) & 
      (1.5--4.1) & (2.1--3.5) & (1.5--3.1) & \\
SMMJ16403+46437 & 7       &  2.2$\pm^{0.8}_{0.6}$ &  
                             2.3$\pm^{0.7}_{0.8}$ & 
                             2.6$\pm^{0.4}_{1.1}$ & 
			     2.6$\pm^{0.4}_{1.0}$ & 
			     2.6$\pm0.5$ & 
			     2.5$\pm0.5$ & 
$A=3.2$, 1.4GHz high, AGN? \\
  & & (1.0--3.3) & (1.1--3.5) & (1.0--3.3) &
      (1.2--3.5) & (1.8--3.5) & (1.5--3.0) & \\
MMJ154127+6615    & 8     & 2.8$\pm^{1.7}_{0.5}$ &  
                            3.6$\pm^{0.9}_{1.4}$ & 
                            2.8$\pm^{1.7}_{0.3}$ & 
			    3.7$\pm^{0.8}_{1.2}$ & 
			    2.7$\pm0.3$ & 
			    2.8$\pm0.3$ & $A=1$ \\
  & & (2.0--6.0) & (2.0--5.9) & (2.0--5.0) &
      (2.0--5.0) & (2.0--3.1) & (2.0--4.1) & \\
MMJ154127+6616    & 8     &  2.8$\pm^{1.7}_{0.3}$ & 
                             2.8$\pm^{2.0}_{0.3}$ &
                             2.8$\pm^{1.8}_{0.3}$ & 
			     2.8$\pm^{2.2}_{0.3}$ & 
			     3.0$\pm^{0.5}_{0.4}$ & 
			     2.9$\pm^{1.0}_{0.4}$ &  $A=1$ \\
  & & (2.0--6.0) & (2.0--6.0)  & (2.5--5.0) &
      (2.1--5.0) & (2.2--3.5) & (2.6--5.3) & \\
MMJ120546$-$0741.5   & 9  &  4.7$\pm^{1.3}_{1.9}$ &
                             3.4$\pm^{2.7}_{1.9}$ & 
                             3.8$\pm^{2.1}_{0.3}$ & 
                             3.8$\pm^{2.0}_{0.3}$ & 
			     3.0$\pm^{0.2}_{0.5}$ & 
			     3.5$\pm1.0$ & \\
  & & (2.5--7.8) & (2.5--7.0) & (3.0--6.0) &
      (2.5--6.0) & (2.5--3.4) & (2.5--6.0) & \\
MMJ120539$-$0745.4   & 9  & 2.8$\pm^{1.7}_{0.6}$ & 
                            2.8$\pm^{1.7}_{0.8}$ & 
			    2.8$\pm^{1.7}_{0.3}$ & 
			    2.5$\pm^{1.7}_{0.5}$ & 
			    2.5$\pm0.3$ & 
			    2.8$\pm^{0.2}_{0.6}$ & \\
  & & (2.0--6.3) & (2.0--6.0) & (2.0--4.9) &
      (2.0--5.3) & (2.0--3.3) & (2.0--3.7) & \\
N1$-$008          & 10    &   0.0$\pm^{0.5}_{0.0}$ & 
                              0.0$\pm^{0.5}_{0.0}$ & 
                              0.0$\pm^{0.4}_{0.0}$ & 
                              0.0$\pm^{0.4}_{0.0}$ & 
			      0.0$\pm^{0.4}_{0.0}$ &
                              0.0$\pm^{0.4}_{0.0}$ & \\
  & & (0.0--0.5) & (0.0--0.5) & (0.0--0.5) &
      (0.0--0.5) & (0.0--0.5) & (0.0--0.5) & \\
\hline
\end{tabular}
\end{center}
\end{minipage}
\end{table*}

\begin{table*}
 \begin{minipage}{140mm}
\begin{center}
\caption{Sources detected at 850$\mu$m or 1.2mm at a $\geq
3.5\sigma$ level, and undetected at any other wavelength. The columns are
identical to those described in Table\,1. The reference
codes are as follows: 
        (1) Scott et al. 2002,
        (2) Eales et al. 2000, 
        (3) Webb et al. 2002,
        (4) Smail et al. 1999,
        (5) Chapman et al. 2002a,
        (6) Dannerbauer et al. 2002.
        (7) Ivison et al. 2002
} \label{2filters}
\begin{tabular}{llccccccl}
\hline
%  object & ref & $z_{\rm phot}^{\rm le1}$ & 68\% CL & 90\% CL & notes \\
  object & ref & $z_{\rm phot}^{\rm le1}$ & $z_{\rm phot}^{\rm
    le1L13}$ & $z_{\rm
    phot}^{\rm le2}$ & $z_{\rm phot}^{\rm le2L13}$ & $z_{\rm
    phot}^{\rm lde1}$ &
    $z_{\rm phot}^{\rm lde2}$ & notes \\
\hline
LH850.2         & 1,7   &  5.9$\pm^{1.3}_{2.4}$ &  
                           5.5$\pm^{1.5}_{1.9}$ & 
                           5.8$\pm^{0.2}_{1.7}$  &
			   5.1$\pm^{0.4}_{1.5}$ & 
			   3.5$\pm^{0.0}_{0.5}$ & 
			   4.3$\pm^{1.2}_{0.8}$ & \\
 & & (3.5--9.0) & (3.3--8.5) & (3.3--6.0) &
     (3.3--6.0) & (2.7--3.5) & (3.2--6.0) & \\
LH850.4         & 1,7   & 5.5$\pm^{2.0}_{1.8}$ &  
                          5.3$\pm^{2.2}_{1.4}$ & 
                          4.7$\pm^{1.3}_{0.5}$ &
			  4.6$\pm^{1.4}_{0.5}$ &
			  3.5$\pm^{0.0}_{0.6}$ & 
			  3.9$\pm^{1.7}_{0.4}$ & \\
 & & (3.0--8.8) & (3.0--8.5) & (3.5--6.0) &
     (3.3--6.0) & (2.6--3.5) & (3.0--6.0) & \\
LH850.5         & 1,7   & 5.3$\pm^{2.2}_{1.3}$ &  
                          6.1$\pm^{1.5}_{1.7}$ & 
                          4.7$\pm^{1.3}_{0.5}$ &
			  6.0$\pm^{0.0}_{1.9}$ & 
			  3.5$\pm^{0.0}_{0.6}$ & 
			  4.5$\pm{1.0}$ & \\
 & & (3.3--9.0) & (3.0--8.5) &  (3.4--6.0) & 
     (3.4--6.0) & (2.6--3.5) & (3.0--6.0) & \\
LH850.13         & 1,7  & 4.9$\pm^{2.6}_{1.4}$ &   
                          5.1$\pm^{2.1}_{1.6}$ & 
                          5.0$\pm1.0$  & 
			  5.0$\pm1.0$ &
			  3.5$\pm^{0.0}_{0.6}$ & 
			  4.6$\pm{1.2}$ & \\
 & & (3.0--8.8) & (3.0--8.4) & (3.0--6.0) & 
     (3.3--6.0) & (2.6--3.5) & (2.7--6.0) & \\
N2850.3         & 1,7  & 4.8$\pm^{2.8}_{0.8}$ & 
                         4.6$\pm^{2.9}_{0.6}$ &
                         5.8$\pm^{0.2}_{1.4}$ &
			 4.8$\pm0.9$ & 
			 3.5$\pm^{0.0}_{0.5}$ & 
			 4.3$\pm{1.0}$ & \\
 & & (3.5--9.0) & (3.5--8.8) & (3.6--6.0) &
     (3.6--6.0) & (2.7--3.5) & (3.2--6.0) & \\
N2850.5         & 1,7  & 5.0$\pm2.0$ &  
                         4.7$\pm^{2.3}_{1.4}$ & 
                         5.0$\pm^{0.7}_{1.5}$ & 
			 4.5$\pm^{1.5}_{0.6}$ & 
			 3.5$\pm^{0.0}_{0.7}$ & 
			 2.5$\pm^{2.0}_{0.5}$ & \\
 & & (2.5--8.6) & (2.7--8.5) & (2.9--6.0) &
     (2.9--6.0) & (2.5--3.5) & (3.0--6.0) & \\
N2850.6         & 1,7  &  5.0$\pm2.0$ &  
                          4.8$\pm^{1.9}_{2.3}$ & 
                          4.8$\pm^{0.8}_{1.3}$ &
			  4.8$\pm1.2$ &
			  3.5$\pm^{0.0}_{0.8}$ & 
			  2.5$\pm^{1.9}_{0.5}$ & \\
 & & (2.5--8.7) & (2.5--8.3) & (3.0--6.0) &
     (2.5--6.0) & (2.5--3.5) & (2.0--5.6) & \\
N2850.8         & 1,7  & 4.0$\pm^{2.7}_{1.5}$ & 
                         4.7$\pm^{2.0}_{2.2}$ & 
                         4.5$\pm^{1.0}_{1.5}$ &
			 4.8$\pm{1.2}$ & 
			 3.0$\pm^{0.5}_{0.3}$ & 
			 2.5$\pm^{0.3}_{1.0}$ & \\
 & & (2.1--8.5) & (2.0--8.4) & (2.6--6.0) &
     (2.6--6.0) & (2.3--3.5) & (2.5--6.0) & \\
N2850.9         & 1,7  & 5.2$\pm^{2.0}_{1.7}$ &  
                         5.0$\pm^{2.0}_{1.8}$ & 
                         5.9$\pm^{0.1}_{1.8}$ &
			 5.0$\pm1.0$ & 
			 3.5$\pm^{0.0}_{0.6}$ & 
			 3.7$\pm^{1.5}_{0.7}$ & \\
 & & (3.0--9.0) & (2.8--8.5) & (3.2--6.0)  &
     (3.1--6.0) & (2.6--3.5) & (2.5--6.0) & \\
N2850.10         & 1,7  & 4.7$\pm^{2.3}_{1.9}$ & 
                          4.5$\pm^{2.5}_{1.7}$ & 
                          4.7$\pm^{0.8}_{1.6}$ &
			  4.8$\pm{1.2}$ & 
			  3.5$\pm^{0.0}_{0.8}$ & 
			  2.5$\pm^{0.5}_{1.0}$ & \\
 & & (2.1--8.5) & (2.1--8.5) & (2.6--6.0) &
     (2.6--6.0) & (2.3--3.5) & (1.5--4.8) & \\
N2850.11         & 1,7  & 5.5$\pm1.9$  &  
                          4.6$^{2.9}_{0.9}$& 
                          5.0$\pm1.0$  & 
			  4.8$^{1.2}_{0.8}$ &
			  3.5$\pm^{0.0}_{0.7}$ & 
			  4.1$\pm{1.2}$ & \\
 & & (3.0--8.9) & (3.0--8.7) & (3.2--6.0)  & 
     (3.2--6.0) & (2.6--3.5) & (2.6--6.0) & \\
CUDSS14.2           & 2 & 5.2$\pm^{2.4}_{1.7}$  & 
                          5.0$\pm^{2.5}_{1.6}$ & 
                          4.7$\pm^{0.8}_{1.0}$  & 
			  5.0$\pm^{0.7}_{1.5}$ & 
			  3.5$\pm^{0.0}_{0.7}$ & 
			  2.5$\pm^{1.2}_{1.0}$ & \\
 & & (2.5--8.7) & (2.5--8.8) & (2.9--6.0) & 
     (2.8--6.0) & (2.5--3.5) & (1.5--5.1) & \\
CUDSS14.4           & 2 & 5.0$\pm^{2.6}_{1.5}$ &    
                          5.1$\pm^{2.5}_{1.6}$ & 
                          4.6$\pm^{1.4}_{0.8}$ & 
			  4.7$\pm^{1.3}_{0.9}$ &
			  3.5$\pm^{0.0}_{0.8}$ &
			  2.5$\pm^{1.2}_{1.0}$ & \\
 & & (2.5--8.8) & (2.5--8.8) & (2.9--6.0) & 
     (2.8--6.0) & (2.4--3.5) & (1.5--5.2) & \\
CUDSS14.5           & 2 & 4.8$\pm^{2.6}_{1.3}$ &    
                          4.9$\pm^{2.7}_{1.4}$ & 
                          4.8$\pm^{1.2}_{0.8}$ &
			  5.3$\pm^{0.7}_{1.2}$ & 
			  3.5$\pm^{0.0}_{0.8}$ & 
			  2.5$\pm^{1.9}_{0.5}$ & \\
 & & (2.8--9.0) & (2.8--9.0) & (3.0--6.0) & 
     (3.0--6.0) & (2.5--3.5) & (2.0--5.7) & \\
CUDSS14.6           & 2 & 4.7$\pm^{2.8}_{1.5}$ &    
                          4.7$\pm^{3.0}_{1.2}$ & 
                          5.2$\pm^{0.8}_{1.3}$ &
			  5.2$\pm^{0.8}_{1.3}$ &
			  3.5$\pm^{0.0}_{0.8}$ & 
			  2.5$\pm^{1.4}_{1.0}$ & \\
 & & (2.5--8.8) & (2.5--8.8) & (2.9--6.0) &
     (2.9--6.0) & (2.5--3.5) & (1.5--5.3) & \\
CUDSS14.8           & 2 & 4.7$\pm^{2.6}_{1.7}$ &    
                          4.7$\pm^{2.0}_{2.2}$ & 
                          4.5$\pm^{1.3}_{1.0}$ &
			  4.7$\pm^{1.9}_{1.2}$ & 
			  3.5$\pm^{0.0}_{0.8}$ & 
			  2.5$\pm^{0.8}_{1.0}$ & \\
 & & (2.5--9.0) & (2.5--9.0) & (2.6--6.0) &
     (2.6--6.0) & (2.3--3.5) & (1.5--5.0) & \\
CUDSS14.10            & 2 & 4.7$\pm^{2.6}_{1.7}$ &    
                            4.8$\pm^{2.5}_{1.8}$ & 
                            3.9$\pm^{2.0}_{0.4}$  & 
			    4.3$\pm1.2$ & 
			    2.9$\pm0.5$ & 
			    2.5$\pm^{0.5}_{1.0}$ & \\
 & & (2.5--9.0) & (2.5--9.0) & (2.6--6.0) & 
     (2.6--6.0) & (2.3--3.5) & (1.5--4.8) & \\
CUDSS3.2              & 3 & 4.5$\pm^{1.8}_{2.2}$ &    
                            3.8$\pm^{2.7}_{1.6}$ & 
                            3.5$\pm^{1.7}_{1.0}$ &
			    3.9$\pm1.4$ & 
			    2.9$\pm{0.5}$ & 
			    2.5$\pm^{0.0}_{0.9}$ & \\
 & & (2.0--8.5) & (1.9--8.5) & (2.0--5.8) &
     (2.2--6.0) & (2.2--3.5) & (1.0--4.0) & \\
CUDSS3.3             & 3 & 3.8$\pm^{2.5}_{1.5}$  &    
                           3.7$\pm^{2.4}_{1.7}$  & 
                           4.0$\pm^{1.2}_{1.5}$  & 
			   4.0$\pm^{1.3}_{1.5}$  & 
			   2.9$\pm0.4$  & 
			   2.5$\pm^{0.0}_{0.9}$ & \\
 & & (2.0--8.4) & (2.0--8.4) & (2.2--6.0) & 
     (2.0--5.8) & (2.1--3.5) & (1.0--4.0) & \\
CUDSS3.4              & 3  & 3.2$\pm^{2.9}_{1.2}$ &   
                             3.7$\pm^{2.8}_{1.7}$ & 
                             4.3$\pm^{0.9}_{1.2}$ &
			     3.9$\pm1.4$ &
			     2.9$\pm0.4$ & 
			     2.5$\pm^{0.0}_{1.0}$ & \\
 & & (2.0--8.5) & (2.0--8.3) & (2.3--6.0) &
     (2.2--6.0) & (2.2--3.5) & (1.0--4.0) & \\
CUDSS3.5            & 3  & 3.7$\pm^{2.6}_{1.7}$ &   
                           3.7$\pm^{2.6}_{1.4}$ & 
                           3.8$\pm^{1.5}_{1.3}$ &
			   3.9$\pm1.4$ &
			   2.9$\pm0.5$ & 
			   2.5$\pm^{0.0}_{1.0}$ & \\
 & & (1.8--8.5) & (1.8--8.5) & (2.0--5.9) &
     (2.0--5.8) & (2.1--3.5) & (1.0--3.9) & \\
CUDSS3.6              & 3 & 3.7$\pm^{2.6}_{1.7}$  &    
                            4.0$\pm^{2.4}_{2.0}$  & 
                            3.7$\pm^{1.5}_{1.2}$  &
			    3.9$\pm1.4$  &
			    2.9$\pm0.4$ & 
			    2.5$\pm^{0.0}_{1.0}$ & \\
 & & (2.0--8.7) & (2.0--8.8) & (2.0--5.8) &
     (2.0--5.8) & (2.2--3.5) & (1.0--3.9) & \\
CUDSS3.7             & 3 & 3.7$\pm^{2.6}_{1.2}$ &    
                           3.7$\pm^{2.6}_{1.2}$ & 
                           4.0$\pm^{1.5}_{1.2}$ & 
			   3.7$\pm{1.3}$ & 
			   2.8$\pm^{0.5}_{0.3}$ & 
			   2.5$\pm^{0.0}_{1.0}$ & \\
 & & (2.0--8.4) & (2.0--8.1) & (2.3--6.0) &
     (2.0--5.7) & (2.0--3.5) & (1.4--4.5) & \\
CUDSS3.11          & 3 & 3.8$\pm^{3.0}_{1.3}$  &  
                         4.5$\pm^{1.7}_{2.5}$ & 
                         3.8$\pm^{1.7}_{1.0}$  & 
			 3.7$\pm^{1.8}_{0.9}$ &
			 2.9$\pm0.4$ & 
			 2.5$\pm^{0.0}_{1.0}$ & \\
 & & (1.9--8.5) & (2.0--8.5) & (2.3--6.0) &
     (2.0--5.7) & (2.2--3.5) & (1.0--4.1) & \\
CUDSS3.12           & 3 & 3.7$\pm^{2.6}_{1.7}$  &  
                          3.2$\pm^{2.1}_{1.2}$  & 
                          3.5$\pm^{1.8}_{1.0}$  & 
			  3.9$\pm1.3$  &
			  2.9$\pm0.5$ & 
			  2.5$\pm^{0.0}_{1.0}$ & \\
 & & (1.9--8.5) & (1.9--8.5) & (2.0--5.8) &
     (2.0--5.8) & (2.2--3.5) & (1.0--3.9) & \\
CUDSS3.13           & 3 & 4.0$\pm^{2.3}_{2.0}$ &  
                          3.2$\pm^{2.2}_{1.5}$ & 
                          3.5$\pm^{1.8}_{1.0}$  & 
			  3.9$\pm1.4$ & 
			  2.9$\pm0.5$ & 
			  2.5$\pm^{0.0}_{1.0}$ & \\
 & & (1.8--8.5) & (1.7--8.5)  & (2.0--5.8) & 
     (2.1--6.0) & (2.1--3.5) & (1.0--4.0) & \\
CUDSS3.16             & 3 & 3.5$\pm^{2.9}_{1.5}$  &  
                            3.2$\pm^{2.2}_{1.2}$  & 
                            3.7$\pm^{1.5}_{1.2}$  & 
			    3.5$\pm^{1.9}_{1.0}$  & 
			    2.9$\pm0.4$ & 
			    2.5$\pm^{0.0}_{1.0}$ & \\
 & & (1.6--8.5) & (1.6--8.5) & (2.1--6.0) & 
     (2.0--5.9) & (2.2--3.5) & (1.0--3.8) & \\
SMMJ02400$-$0134       & 4 & 5.0$\pm^{2.6}_{1.5}$ &  
                             5.2$\pm^{2.3}_{1.7}$ & 
                             4.9$\pm^{1.1}_{0.9}$ & 
			     5.2$\pm^{0.8}_{1.2}$ & 
			     3.5$\pm^{0.0}_{0.7}$ & 
			     4.0$\pm^{0.7}_{2.0}$ &  
 $A>1.9$ (2.5?) \\
 & &  (3.0--9.1) & (3.0--9.1) & (3.1--6.0) & 
      (3.1--6.0) & (2.5--3.5) & (2.0--5.5) & \\
SMMJ04431+0210 (N4)  & 4 & 4.5$\pm^{2.5}_{1.6}$ &  
                           4.6$\pm^{2.2}_{1.6}$ & 
                           4.5$\pm^{1.4}_{1.0}$ & 
			   4.2$\pm^{1.6}_{0.7}$ & 
			   3.5$\pm^{0.0}_{0.7}$ & 
			   2.4$\pm1.0$ & 
 $A>1.$\\
 & & (2.5--8.8) & (2.5--8.7) & (2.7--6.0) & 
     (2.7--6.0) & (2.5--3.5) & (1.5--4.9) & \\
SMMJ04541$-$0302  & 5  & 4.7$\pm^{2.0}_{2.2}$  &  
                         4.0$\pm^{2.6}_{1.5}$ & 
                         4.7$\pm1.2$  &
			 4.7$\pm^{0.8}_{1.6}$ & 
			 3.0$\pm^{0.5}_{0.3}$ & 
			 2.5$\pm^{0.6}_{1.0}$ &
 $A=2.6$ \\
 & & (2.2--8.5) & (2.3--8.5) & (2.6--6.0) &
     (2.6--6.0) & (2.4--3.5) & (1.5--4.7) & \\
\hline
\end{tabular}
\end{center}
\end{minipage}
\end{table*}

\begin{table*}
 \begin{minipage}{140mm}
 \setcounter{table}{2}
\begin{center}
\caption{(cont.)} 
\label{2filters}
\begin{tabular}{llccccccl}
\hline
%  object & ref & $z_{\rm phot}^{\rm le1}$ & 68\% CL & 90\% CL & notes \\
  object & ref & $z_{\rm phot}^{\rm le1}$ & $z_{\rm phot}^{\rm
    le1L13}$ & $z_{\rm
    phot}^{\rm le2}$ & $z_{\rm phot}^{\rm le2L13}$ & $z_{\rm
    phot}^{\rm lde1}$ &
    $z_{\rm phot}^{\rm lde2}$ & notes \\
\hline
SMMJ04542$-$0301  & 5   & 4.7$\pm2.2$ &  
                          4.3$\pm^{2.7}_{1.5}$ & 
                          4.2$\pm^{1.7}_{0.7}$ & 
			  4.5$\pm^{1.4}_{1.0}$ & 
			  3.3$\pm^{0.2}_{0.6}$ &
			  2.5$\pm^{0.9}_{1.0}$ &  
$A=4.5$  \\
 & & (2.5--8.9) & (2.1--8.5) & (2.7--6.0) & 
     (2.6--6.0) & (2.4--3.5) & (1.5--4.9) & \\
SMMJ04543$+$0256  & 5   &  3.9$\pm^{2.7}_{1.4}$ &  
                           3.8$\pm^{2.7}_{1.6}$ & 
                           4.0$\pm^{1.5}_{1.1}$ &
			   4.0$\pm^{1.5}_{1.1}$ & 
			   2.9$\pm0.4$ & 
			   2.5$\pm^{0.0}_{1.0}$ &  
$A=3.1$ \\
 & & (2.0--8.4) & (2.0--8.5) & (2.4--6.0) & 
     (2.0--6.0) & (2.0--3.5) & (1.5--4.5) & \\
SMMJ04543$+$0257  & 5   &  4.7$\pm^{2.2}_{1.7}$ &  
                           4.5$\pm^{2.1}_{1.5}$ & 
                           4.2$\pm^{1.3}_{1.1}$ &
			   4.5$\pm^{1.0}_{1.4}$ & 
			   3.5$\pm^{0.0}_{0.8}$ & 
			   2.5$\pm^{0.7}_{1.0}$ &  
$A=3.4$ \\
 & & (2.0--8.2) & (2.5--8.3) & (2.6--6.0) & 
     (2.6--6.0) & (2.3--3.5) & (1.5--4.8) & \\
SMMJ10237$+$0412  & 5   &  4.0$\pm^{2.9}_{1.5}$  &  
                           3.9$\pm^{3.0}_{1.4}$  & 
                           4.5$\pm^{0.7}_{2.0}$  & 
			   4.7$\pm^{0.8}_{1.8}$  & 
			   2.9$\pm0.4$  &
			   2.5$\pm^{0.0}_{1.0}$ &  
$A=3.5$ \\
 & & (2.0--8.7) & (2.0--8.7) & (2.0--6.0) &
     (2.6--6.0) & (2.2--3.5) & (1.0--4.1) & \\
SMMJ14573$+$2220  & 5   &  4.0$\pm^{2.7}_{1.5}$ &  
                           4.2$\pm^{2.6}_{1.7}$ & 
                           4.0$\pm^{1.6}_{1.0}$ &
			   4.2$\pm^{1.8}_{0.8}$ & 
			   2.9$\pm0.4$ & 
			   2.5$\pm^{0.1}_{1.0}$ &  
 $A=2.8$ \\
 & & (2.0--8.5) & (2.0--8.6) & (2.0--6.0) & 
     (2.5--6.0) & (2.2--3.5) & (1.5--4.7) & \\
SMMJ17223$+$3207  & 5   &  4.0$\pm^{2.5}_{1.5}$ &  
                           4.2$\pm^{2.6}_{1.7}$ & 
                           3.9$\pm^{1.6}_{1.0}$ &
                           4.0$\pm^{1.1}_{1.5}$ & 
			   3.0$\pm^{0.3}_{0.5}$ & 
			   2.5$\pm^{0.1}_{1.1}$ &  
$A=2.8$ \\
 & & (2.0--8.4) & (2.0--8.4) & (2.4--6.0) & 
     (2.5--6.0) & (2.2--3.5) & (1.5--4.6) & \\
SMMJ22471$-$0206     & 4 &  4.8$\pm^{2.4}_{1.8}$ &
                            4.8$\pm^{2.3}_{1.8}$ & 
                            4.7$\pm^{1.3}_{0.9}$ & 
			    4.9$\pm{1.1}$ & 
			    3.4$\pm^{0.1}_{0.6}$ & 
			    2.9$\pm^{1.3}_{1.0}$ &  
$A>1.9$ (2.5?)\\
 & & (2.5--8.8) & (2.5--8.8) & (2.8--6.0) &
     (2.8--6.0) & (2.5--3.5) & (1.5--5.2) &  \\
MMJ120517$-$0743.1 & 6 & 7.0$\pm^{1.5}_{2.4}$ &  
                         7.1$\pm^{1.4}_{2.6}$ & 
                         6.0$\pm^{0.0}_{1.6}$ & 
			 6.0$\pm^{0.0}_{1.6}$ & 
			 3.5$\pm^{0.0}_{0.4}$ & 
			 4.4$\pm^{1.4}_{0.4}$ &  \\
 & & (4.3--10.) & (4.3--10.) & (3.6--6.0) & 
     (3.5--6.0) & (2.7--3.5) & (3.5--6.0) & \\
W$-$MM11              & 5 &  4.2$\pm^{2.0}_{2.2}$ &  
                             3.4$\pm^{2.9}_{1.5}$ & 
                             3.7$\pm^{1.6}_{1.2}$ & 
			     3.9$\pm^{1.6}_{1.2}$ & 
			     2.9$\pm0.4$ &
			     2.5$\pm^{0.0}_{0.9}$ &
$z_{\rm spec}=2.98$ \\
 & & (2.0--8.7) & (2.0--8.6) & (2.0--5.8) & 
     (2.2--6.0) & (2.0--3.5) & (1.0--4.0) & \\

\hline
\end{tabular}
\end{center}
\end{minipage}
\end{table*}

 \subsubsection{Notes on individual sources}

%check references CY00 or YC02, abbreviate YC,CY

\hspace{0.7cm}{\bf LH850.1:}
Other redshift estimates are 
$2.95\pm^{1.49}_{0.98}$
based on the 1.4GHz/850$\mu$m spectral index technique; and
$2.72\pm 0.37$ based on
a $\chi^2$ minimization with one template SED (YC02) .
Our own estimate $z=2.6\pm^{0.4}_{0.5}$  (le2) 
is consistent with these measurements.

{\bf LH850.12:}
No SED can reproduce the observed photometry of  this source. Ivison
et al. (2002) find evidence of variability at 1.4GHz, and report
associated X ray emission. They conclude
that LH850.12 might be a radio-loud QSO. This kind of objects is actually
not represented among our SEDs, and thus the redshift we derive is unreliable.
Note that there are no counterpart mock-galaxies in the vicinity of
the error box of its $C-C-z$ diagram in Fig.~A1.

{\bf HDF850.1:} Other redshift estimates are $>2.6$,
based on the 1.4GHz/850$\mu$m spectral index technique; and
$4.11\pm 0.51$  based on
a $\chi^2$ minimization with one template SED (YC02). Our own estimate
$z=4.1\pm^{0.6}_{0.5}$ (le2) 
is in good agreement with these measurements, and is
unaffected by 
the inclusion or exclusion 
of the possible lensing amplification experienced by 
this source ($A\approx 3$, Dunlop et al. 2002).

{\bf CUDSS14.1:}~~Other published redshift estimates are $2.01
\pm ^{1.10} _{0.71}$, based on the
1.4GHz/850$\mu$m spectral index technique; and $2.06\pm 0.31$ 
 based on a $\chi^2$ minimization with one template
SED (YC02). Both these estimates are in the low-redshift tail of our redshift
distribution, which at a 68\% significance level places this object at
$z=3.8\pm ^{0.7} _{0.8}$ (le2).

{\bf CUDSS14.18:} The inclusion of the 450 and 850$\mu$m fluxes as detections
at 2.3 and 2.8$\sigma$ (instead of upper limits) gives a similar distribution
to the one shown in Figure~A2.  
Our redshift estimate places this object at
$z=0.7\pm^{1.3}_{0.2}$  with a 68\% confidence level (le2), in good agreement
with its true spectroscopic redshift, $z_{\rm spec}=0.66$.
Other redshift estimates are $1.12 \pm ^{0.53}
_{0.45}$, based on the 1.4GHz/850$\mu$m spectral index
technique; and $1.11\pm0.21$  based on a $\chi^2$
minimization with one template SED (YC02).

{\bf SMMJ00266+1708:} Other redshift estimates are $3.49\pm^{2.03}_{1.23}$
based on the 1.4GHz/850$\mu$m spectral index technique; and $3.50\pm0.45$ 
 based on a $\chi^2$ minimization with one template SED (YC02).  Our
estimate $z=2.7\pm^{2.3}_{0.2}$ (le2) is consistent with these measurements. The
redshift distribution is bimodal, due to the range of SEDs used in the
redshift estimation technique.

{\bf SMMJ02399$-$0136:} The complete SED reported in Frayer et al. (2000),
Ivison et al. (1998) is inconsistent with the SEDs of our template library at
any redshift.  This is reported to be an AGN (Smail et al. 1999).
One
possible explanation for the failure to detect SMMJ02399-0136 at 3.4cm
is a high-frequency break (between 5.5cm and 0.9cm) in the synchrotron
spectrum. Such sharp spectral features are not taken into account in
the fitting of the radio spectrum. We note however that excluding the
3.4cm upper-limit from the modelling of the multi-wavelength data
provides a redshift estimate $z=2.8\pm^{1.3}_{0.3}$ (le2) that is in good
agreement with the true redshift of SMMJ02399$-$0136: $z=2.80$.
Other redshift estimates are $1.65\pm^{0.80}_{0.53}$ based on the
1.4GHz/850$\mu$m spectral index technique; and $2.83\pm0.38$ 
based on a $\chi^2$ minimization with one template SED (YC02).

{\bf SMMJ09429$+$4658 (H5):} Other redshift estimates are $\ge3.6$ based on
the 1.4GHz/850$\mu$m spectral index technique; and $3.86\pm0.49$ 
 based on a $\chi^2$ minimization with one template SED (YC02).  Our
estimates actually place this source at a slightly lower redshift
$z=2.4\pm^{1.5}_{0.4}$ (le2). 
Given the sub-mm detection, the low flux density at
1.4\,GHz excludes almost all of the template SEDs in our library.

{\bf SMMJ14009+0252:} Other redshift estimates are
$1.27\pm^{1.59}_{0.54}$ based on the 1.4GHz/850$\mu$m spectral index
technique; and $1.30\pm 0.23$ based a on a $\chi^2$ minimization with
one template SED (YC02).  However large discrepancies ($\sim 3\sigma$)
exist between the template SED of YC02 and the observed 450$\mu$m and
1.4GHz fluxes, suggesting the fit will have a high reduced-$\chi^2$,
and that their errors (inferred for good matches) might be
under-estimated.  The flattening of the spectrum at 450$\mu$m,
relative to the Raleigh-Jeans tail defined by the 850$\mu$m and 1.35mm
fluxes, supports a higher redshift estimate. Our own fit indicates
that this object has a redshift $z\sim 4.1\pm0.8$ (le2). The discrepancies
between different redshift estimators might be attributed to the
possible AGN nature of this object (Ivison et al. 2000).

{\bf SMMJ14011+0252:} 
Other redshift estimates are 
$2.53\pm^{1.24}_{0.82}$
based on the 1.4GHz/850$\mu$m spectral index technique; and
$2.73\pm0.37$  based on
a $\chi^2$ minimization with one template SED (YC02).
Our own estimate is $z=2.7\pm^{0.6}_{0.7}$ (le2) in good agreement with
these estimates, and with the spectroscopic redshift of the optical 
counterpart $z_{\rm spec}=2.57$.

{\bf HR10:} 
Other redshift estimates are 
$1.49\pm^{0.75}_{0.53}$
based on the 1.4GHz/850$\mu$m spectral index technique; and
$1.75\pm 0.28$  based on
a $\chi^2$ minimization with one template SED (YC02).
Our own estimate $z=1.5\pm^{0.9}_{0.0}$ (le2) is in good agreement with
their measurements, and with the true spectroscopic redshift of this
source $z_{\rm spec}=1.44$.

%%%%%%%

{\bf SMMJ02399$-$0134:} Other redshift estimates are
$1.17\pm^{0.53}_{0.45}$ based on the 1.4GHz/850$\mu$m spectral index
technique; and $0.93\pm0.19$ based on a $\chi^2$ minimization with one
template SED (YC02).  The spectroscopic redshift of this source
($z_{\rm spec}=1.06$) is inconsistent with our best estimate
of the redshift, which within a 68\% confidence interval places it at
$z=2.3 \pm 0.8$ (le2). Redshifts in the range
$1.0 \le z\le 1.5$ have a  10\% integrated probability.

{\bf SMMJ16403$+$46437:} 
The detection at 1.4GHz (593$\mu$Jy) is high when compared 
with the 850$\mu$m level, and there is no acceptable
correspondence with any of the SEDs in our template library.
This, plausibly, might be a
misidentification or a radio-quiet AGN, with enhanced radio luminosity
compared to the rest-frame FIR luminosity.

{\bf SMMJ16403$+$46440:} 
Treating the 1.35mm observation as a detection (2.2$\sigma$) provides
no additional constraint and gives a similar redshift 
distribution to the one derived from the 450$\mu$m/850$\mu$m colour
and the 1.4GHz and 1.35mm upper limits,
shown in Figure~A2.

\subsection{Cumulative redshift distribution of the sub-mm galaxy population}

It is straightforward to calculate the cumulative redshift distribution for
the sub-mm galaxy population as the coaddition of the individual probability
distributions. It should be noted that the calculation carried out in this
paper has the advantage of including the whole redshift probability
distribution of each individual galaxy, and not just the mean values of the
distributions.  We have included in this calculation of the cumulative
distribution only those sub-mm sources identified in wide-area 
blank-field surveys (UK
8mJy survey and the CUDSS) at a $\geq 3.5\sigma$ level: 50 sources. 
We do not include sub-mm galaxies identified in surveys
carried out towards lensing clusters, or targeted observations of other
populations of galaxies (radio, ISO, Lyman-break galaxies or extremely
red objects), or catalogs
that have not been fully published (MAMBO surveys), since they can distort and bias
the characteristic redshift distribution of the sub-mm galaxy population.

\begin{figure}
    \vspace*{-2.cm}
    \hspace*{8cm}
    \figl{7cm}{26}{26}{573}{737}{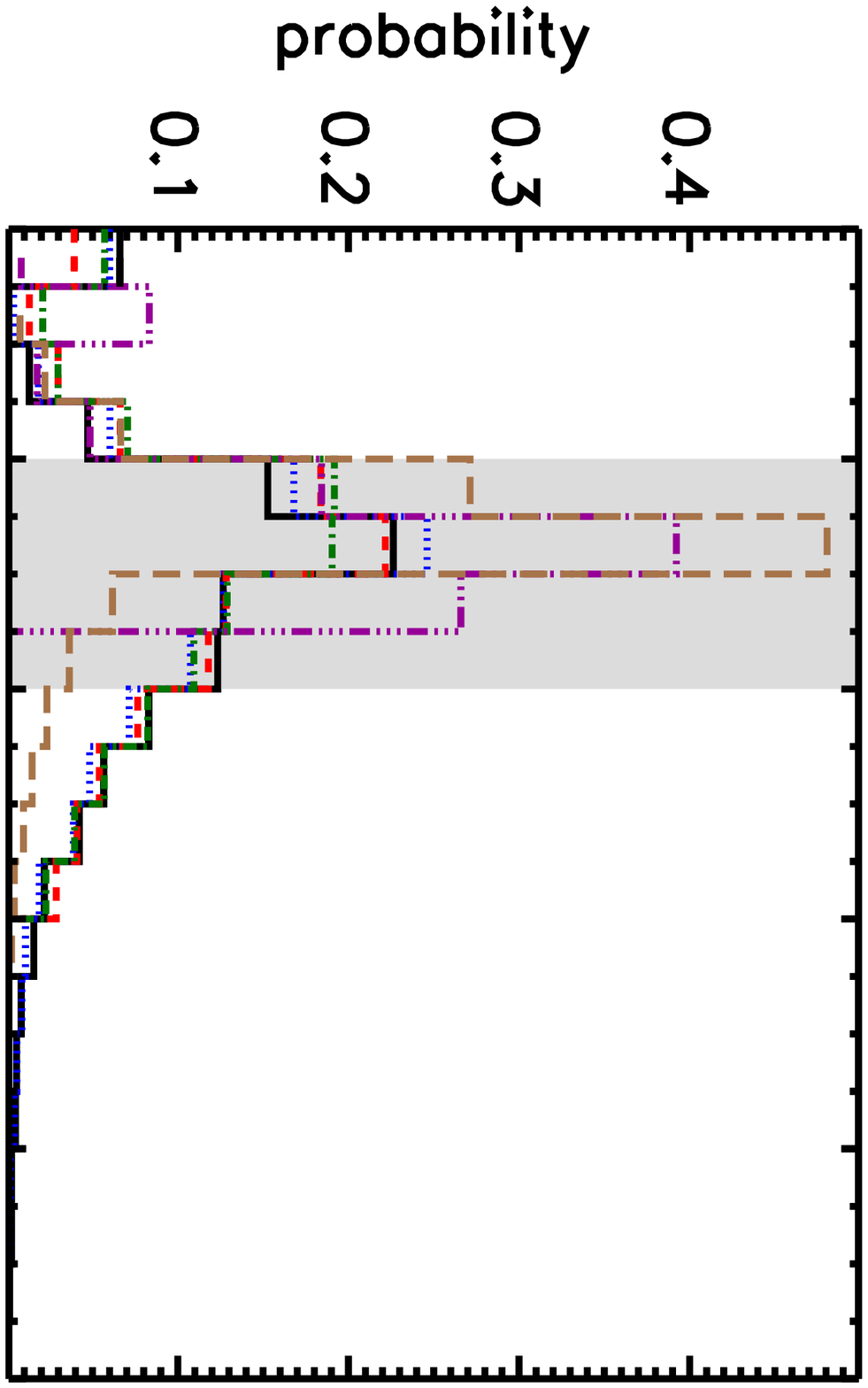}{90} 

    \vspace*{-4.6cm}
    \hspace*{8cm}
    \figl{7cm}{26}{26}{573}{737}{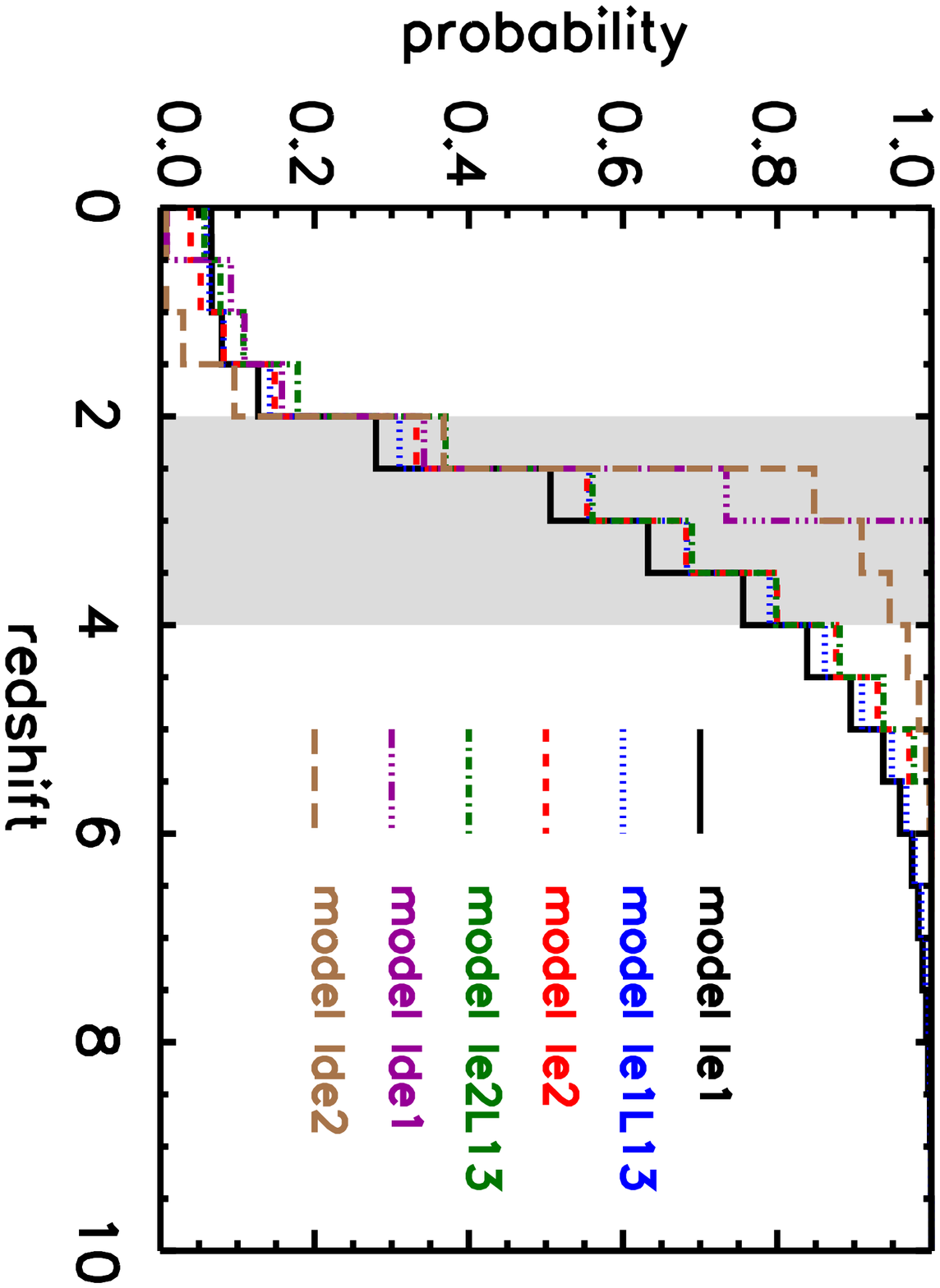}{90} 
    \caption{Combined redshift distribution for the population of
      sub-mm
galaxies in the 8mJy SCUBA survey  with one or more measured colours
(the 14 galaxies contained in Tables~1 \& 2).
The different lines represent different evolutionary models
assumed for the sub-mm galaxy population (section 2).
{\it Upper panel:} discrete bin probabilities.
{\it Lower panel:} cumulative distribution.
}
\label{fig:cumz8mJy}
\end{figure}

\begin{figure}
    \vspace*{-2.cm}
    \hspace*{8cm}
    \figl{7cm}{26}{26}{573}{737}{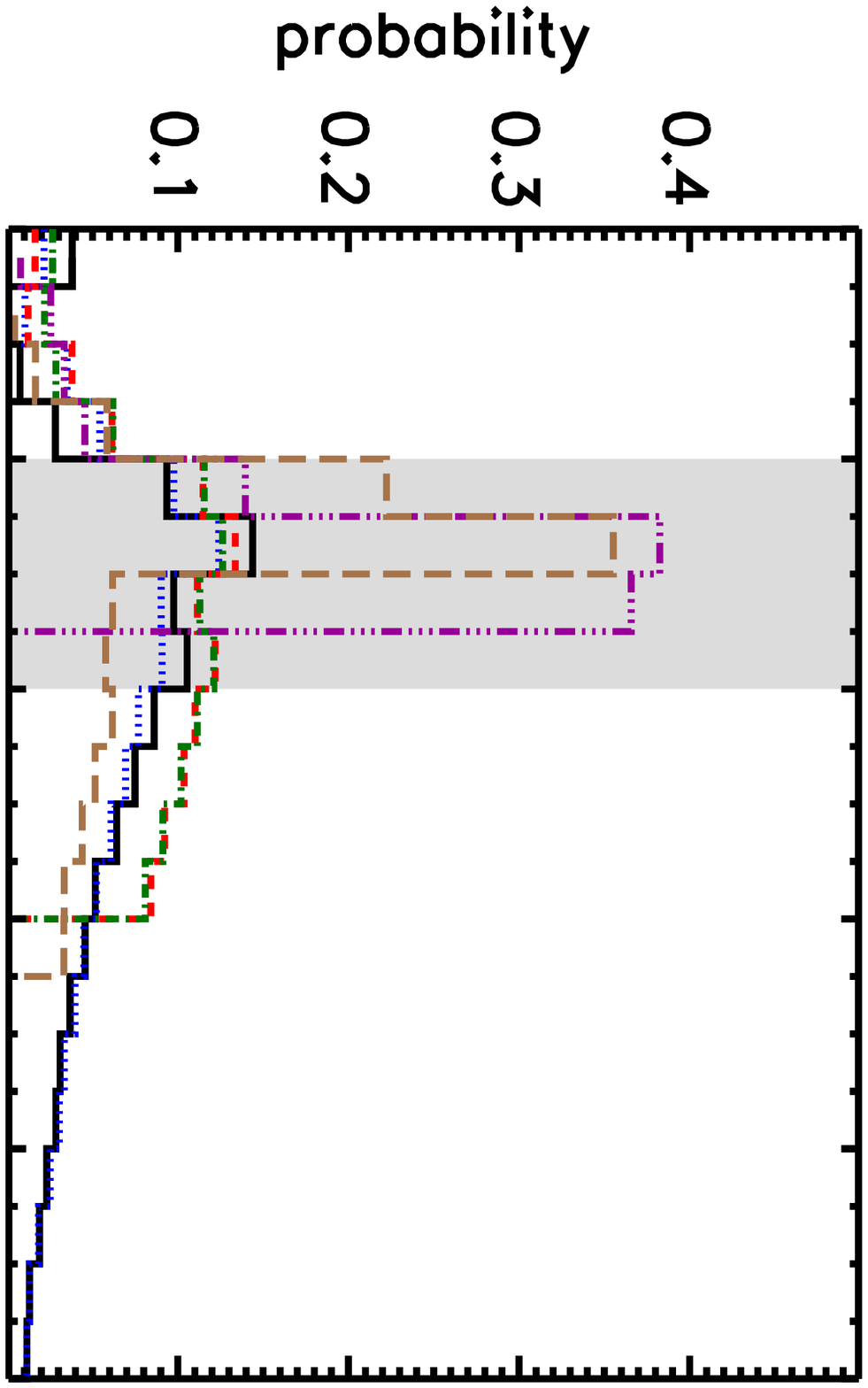}{90} 

    \vspace*{-4.6cm} \hspace*{8cm}
    \figl{7cm}{26}{26}{573}{737}{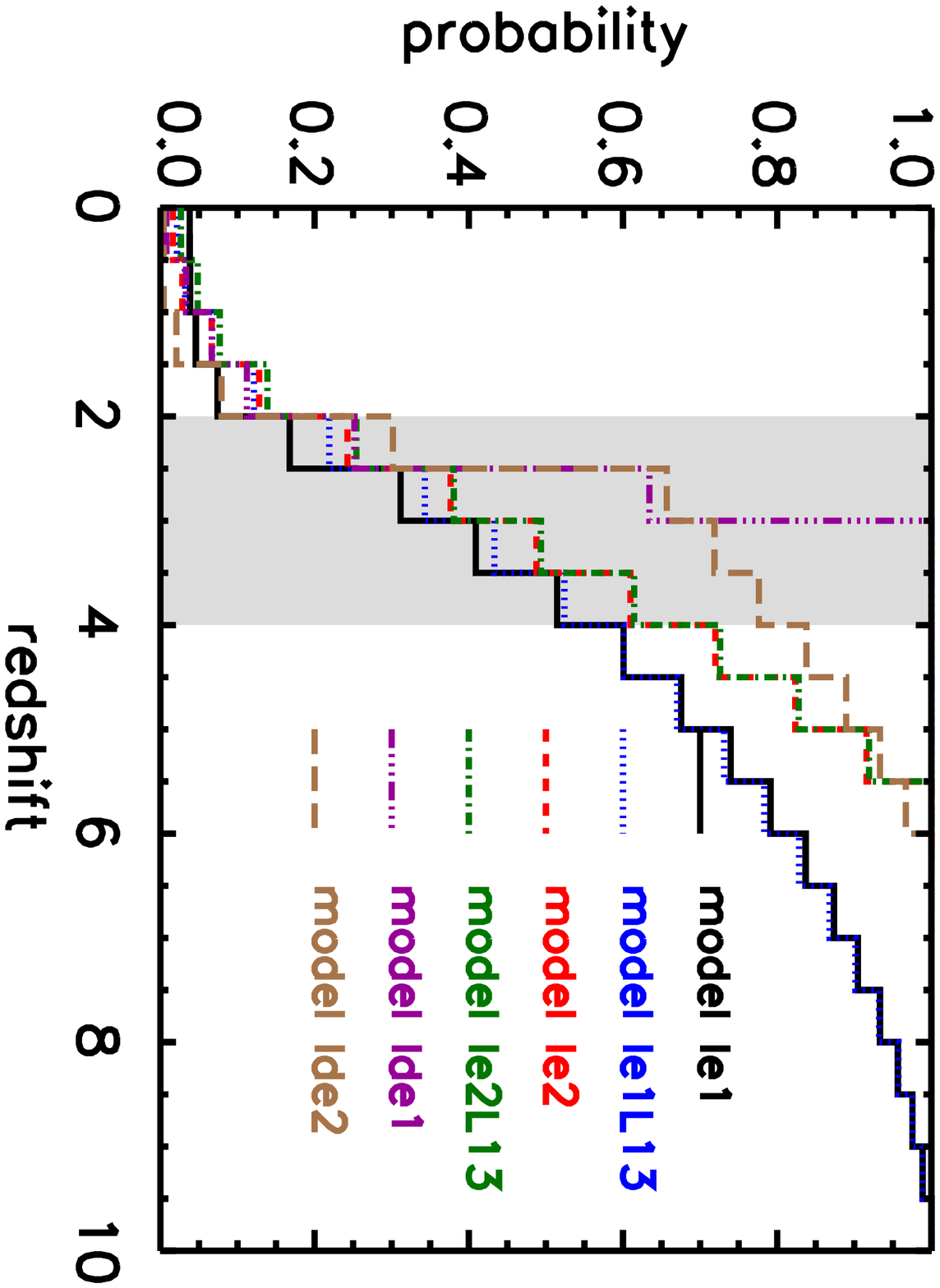}{90} \caption{Combined
    redshift distribution for the population of all 50 sub-mm galaxies
    detected in wide-area surveys at 850$\mu$m ($> 3.5\sigma$) and
    listed in Tables 1,2, and 3. The
    different lines represent different evolutionary models assumed
    for the sub-mm galaxy population (see text).  Beware that the
    extended tail towards high-redshifts is due to sources which
    are just detected at 850$\mu$m. The upper limits at other wavelengths do
    not usually help to constrain their redshifts between $2\le z\le 10$, and
    hence their probability distributions are very flat in those
    regimes.  {\it Upper panel:} discrete bin probabilities.  {\it
    Lower panel:} cumulative distribution.  }
\label{fig:cumz}
\end{figure}

Half of sources in complete flux-limited sub-mm samples, such
as the UK 8mJy and CUDSS surveys, have a single 850$\mu$m detection
with one, or more, additional upper-limits at other wavelengths: 28/50
sources detected ($> 3.5\sigma$) at 850$\mu$m belong to this category.
Since the redshift distributions of these sources are the most
dependent on the priors used, the interpretation of  the
combined redshift distributions should be based on the 
range of results given
by the different evolutionary models.

Fig.~\,\ref{fig:cumz8mJy} shows the cumulative distribution and the
combined redshift distribution of the 14 galaxies in the UK 8mJy survey
with redshift distributions  derived
from SEDs which contain at least 2-band detections (objects in
Tables~1 and 2).  This figure indicates that 65--90 per cent of the sub-mm
galaxies could have redshifts in the range $2 \lsim z \lsim 4$.  The results
are consistent with the expectation 
%prior understanding 
that sub-mm galaxies
detected in the SCUBA (and also MAMBO) surveys represent a high-$z$
population (Dunlop 2001, Blain et al. 2002).

The sources in the CUDSS sample with one or more measured colours (9
sources) appear to have a significantly different lower redshift
distribution, with 40 per cent of the galaxies at $z \lsim 2$ and
approximately 50 per cent of the galaxies at $2 \lsim z \lsim 4$.  The
high redshift tail of the CUDSS sample is similar to that shown by the
8mJy survey with $\sim$10 per cent probability at $z > 4$.  The two
galaxies responsible for the low-redshift difference are CUDSS3.1 and
3.9, both with relatively high 450$\mu$m flux densities $S_{450\mu
{\rm m}} \ge 77$~mJy, compared with the 30--40~mJy detection levels
for the UK 8mJy survey sources.  Taking the data for the CUDSS sample
at face-value suggests that it includes an over-density of low-$z$
galaxies at $z\lsim2$, compared to the galaxies in the UK 8mJy survey,
and vice-versa. This discrepancy, however, could also be explained by
a difference in the flux-calibration of the 450$\mu$m data in the two
surveys.

Fig.~\,\ref{fig:cumz} shows the cumulative redshift distributions of
the sub-mm galaxy population for the 50 galaxies in the 8mJy and CUDSS
surveys detected at a $\geq 3.5\sigma$ level.  This extended sample
includes the 27 sub-mm galaxies with a single detection at $850\mu$m
and one, or more, upper limits at other wavelengths (Table\,3).  The
combined redshift distributions, for the different evolutionary models
under consideration, imply that the majority of the sub-mm galaxies,
approximately 50 -- 90 per cent, lie between $z = 2 - 4$.  The
remainder of the galaxies, $\sim 10$ per cent, lie at $z <
2$. Fig.~\,\ref{fig:cumz} also shows that $\lsim 50$ per cent  of the galaxies
have colours that are consistent with $z > 4$.  The distribution
of the high-redshift tail is very sensitive to the choice of
evolutionary model, since the individual redshift distributions for
the 27 sources with only a single detection are necessarily very flat
over the redshift range $z\gsim 2$ (see Fig.~A3). For these galaxies,
the constraint on their redshift distributions comes mainly from the
flux detected at 850$\mu$m and the ability to reproduce it with the
evolution of the $60\mu$m local luminosity function. More extensive
photometry, especially including shorter sub-mm wavelengths, is
crucial to better constrain these flat distributions, as illustrated
by Fig. 7 (see also Fig.6 in paper~I).

\section{The likelihood of mis-identifying the radio counterparts of 
high-$z$ sub-mm galaxies}

The strength of the correlation between the FIR and radio luminosities
in
 starburst galaxies (Helou et al. 1985, Condon 1992)
has supported the expectation that {\em all}
high-$z$ star forming sub-mm galaxies should also have a detectable
radio-counterpart, provided the radio observations are deep enough.

Although originally encouraged by this assumption, some of the deepest VLA
searches for radio emission associated with SCUBA and MAMBO sources
have met with varying degrees of success.

\begin{figure}
    \vspace*{-2.5cm}
\hspace*{8cm} \figl{7cm}{26}{26}{573}{737}{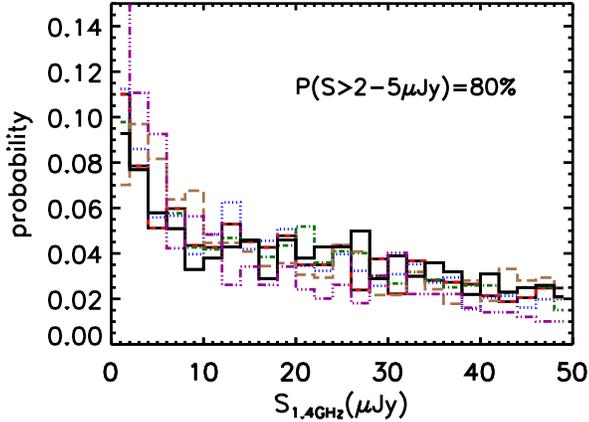}{90}
\caption{Distribution of intrinsic fluxes at 1.4GHz corresponding to a
population of galaxies discovered at 850$\mu$m, at a level $\ge8$mJy.
Depending on the evolutionary model adopted, 80\% of the sources would
have a flux in excess of 2 to 5$\mu$Jy, and hence in order to conduct
a parallel 1.4GHz imaging follow-up program with 80\% completeness,
the 1$\sigma$ depth of the images should be 1 to
2$\mu$Jy.
}
\label{fig:radio_prob}
\end{figure}

The deep VLA surveys of the Lockman Hole and ELAIS N2 fields (Ivison
et al. 2002) have a $1\sigma$ noise limit $\geq 5-9\mu$Jy at 1.4~GHz.
These surveys have identified 11/16 and 8/14 sub-mm sources at
1.4GHz, respectively, with peak radio fluxes above 4$\sigma$ and
integrated fluxes above 3$\sigma$, after rejecting 5 and 1 sub-mm sources
from the parent samples, respectively, because of their high sub-mm
noise levels. From these, there are 9/16 and 5/14 sources in the
Lockman Hole and ELAIS N2 surveys, respectively, with a final  S/N$>$3
when one considers the additional errors due to calibration and fitting 
a gaussian to the peak of the radio emission. If one limits
the analysis to sources $\ge 8$~mJy at 850$\mu$m, then 4/10 and 4/8
sources have been simultaneously detected with S/N$>3$ at 850$\mu$m
and 1.4GHz.  This is in clear contrast to the claims of previous
authors \cite{fox01} that bright sub-mm sources ($>8$mJy at 850$\mu$m)
should be detected at flux densities of $\sim 100 \mu$Jy at 1.4\,GHz.

In order to understand the properties  of the  radio associations, we
  have calculated the probability of detection at 1.4GHz for 
a mock 8mJy survey,
  following the same methodology as in paper~I, but using the
  extended library of SEDs (section 2) . The intrinsic 
850$\mu$m fluxes of the galaxies are convolved  with $1\sigma \sim $ 2.6mJy
  measurement errors, and 9 per cent calibration uncertainty. 
Fig.\ref{fig:radio_prob} shows the distribution of intrinsic 1.4GHz
  fluxes for the detected sources (including errors) above the 8mJy 
threshold at 850$\mu$m. 

We conclude that future surveys at 1.4\,GHz must reach a 3$\sigma$
sensitivity of 2--5$\mu$Jy if radio observations are to provide
counterparts to 80\% of the high-$z$ sub-mm galaxies with 850$\mu$m
flux densities $\ge 8$\,mJy, according to the redshift distributions
assumed in the alternative evolutionary models.  
Thus,  current radio observations have insufficient
sensitivity to detect the bulk of the sub-mm population with
adequate S/N, and we must
wait for the next generation of new and upgraded facilities
(eg. e-VLA, e-MERLIN, SKA).

In  the case of the UK 8mJy survey, 
  Fig.~10 shows that 40--50\% (Lockman Hole) and 50--60\% (ELAIS N2)
  of the sources should have an intrinsic 1.4GHz flux lower than the
  3$\sigma$ threshold, in the absence of any radio-loud
  AGN emission. 
These levels agree well with the number of radio
  counterparts to $\ge 8mJy$ sources (Scott et al. 2002,
  Ivison et al. 2002), bearing in mind the actual variation in
  sensitivity across the SCUBA wide-area survey fields.

This potential for misidentifying the correct radio sources associated
with blank-field sub-mm galaxies will inevitably lead to incorrect
estimates of photometric redshifts if they are derived in part from
the 1.4\,GHz fluxes. In the calculation of the cumulative redshift
distribution (section 3.3) we have treated all of the published
associations of 1.4\,GHz and 850$\mu$m sources as the correct radio
counterparts to the sub-mm galaxies.

Despite the rapid acceptance of the radio--sub-mm flux density ratio,
$\alpha^{1.4}_{345}$ (CY99, CY00), as the preferred method with which
to estimate the redshift distribution of the sub-mm galaxy population,
this diagnostic ratio loses its power to discriminate between
photometric redshifts at $z > 2$ (CY99, CY00).  We have already
demonstrated in paper\,I (Fig.~11) the detrimental effect of the large
dispersion in radio--sub-mm colours of the local template galaxies on
the accuracy of the derived photometric redshifts of the sub-mm blank
field sources. Furthermore, the existing radio interferometers will
remain limited in their ability to map the large areas $\sim$ 1--40
sq. deg. that will be surveyed by the next generation of sub-mm and
FIR experiments (e.g. BLAST, Herschel, SIRTF) down to the necessary
sensitivity levels (3$\sigma \sim 2-5 \mu$Jy at 1.4GHz), making the
use of $\alpha^{345}_{1.4}$ an impractical measure of redshift for the
majority of the population. Deep interferometric radio observations,
however, remain a practical way to pinpoint the source of the sub-mm
emission with sufficient positional accuracy to investigate galaxy
morphologies and exploit optical/IR spectroscopic line diagnostics.

This limitation of the radio--sub-mm technique 
prompted the design and construction of a Balloon-borne Large Aperture
Submillimetre Telescope, BLAST \cite{devlin01}.  The primary
scientific goal of BLAST is to break the current ``redshift deadlock''
by providing photometric redshifts with sufficient accuracy, without
the prior necessity for secure optical, IR or radio counterparts.
BLAST will use rest-frame sub-mm and FIR colours to derive redshifts
with an accuracy $\Delta z\sim \pm 0.5$ for $> 1000$ galaxies in the redshift
range $0 < z < 6$ detected in wide-area ($> 0.5-10$\,deg$^{2}$) surveys
at 250, 350 and 500$\mu$m \cite{hughes02}. 

This redshift
accuracy is sufficient to guide the choice of possible front-end
spectrographs, and their receiver tunings, on large single-dish
ground-based mm and cm telescopes (e.g. 100-m GBT--Green Bank Telescope and
50-m LMT--Large
Millimetre Telescope) in the
effort to detect redshifted CO emission-lines. We conclude this paper
with the description of one application 
of this technique.

\begin{figure}
    \vspace*{-2.5cm}
\hspace*{8cm}
    \figl{7cm}{26}{26}{573}{737}{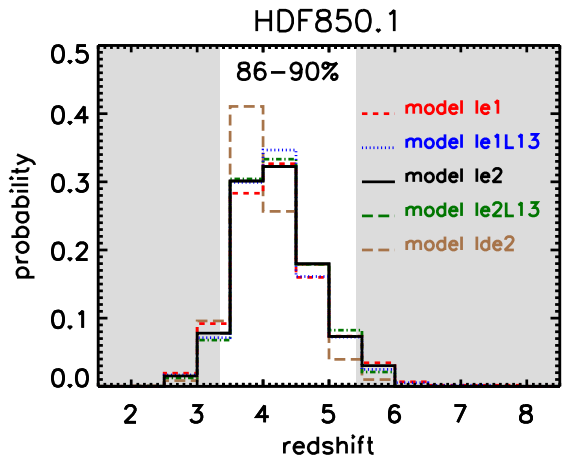}{90} 
    \caption{Redshift probability distributions for HDF850.1. The
different line types and colours correspond to the different
evolutionary models considered. Model led1 is not plotted, since  it
excludes sources at $z>3.5$ due to the strong prior
decline of the density  and luminosity of sources.
The unshaded area corresponds to the redshift range where the
$^{12}$CO ($J=1-0$) line lies in the
observable K-band window (18.5 -- 26.0~GHz), accessible with the 100m GBT. There is
an 86-90\% probability of detecting the J=1-0 CO-line in this GBT window
}  
\label{fig:hdf_co}
\end{figure}

\section{Broad-band mm--cm spectroscopy}

Until recently the brightest sub-mm source in the Hubble Deep Field,
HDF850.1 \cite{hughes98}, has avoided the detection of its X-ray,
optical, IR or radio counterpart, despite an accurate interferometric
position at 236~GHz \cite{downes99} and the unprecedented depth of the
multi-wavelength HDF surveys
\cite{williams,thompson,richards99,hornschmeier}. It is only after the
accurate subtraction of the elliptical galaxy, 3$-$586, at $z \sim 1$,
however, that a faint lensed IR counterpart ($K=23.5$, $H-K = 1.5$, $I-K>
5.1$) to HDF850.1 has been discovered \cite{dunlop02}. 

This IR detection now gives greater confidence to the likelihood that
the coincident radio emission, with a flux density of 16$\pm 4 \mu$Jy
at 1.4\,GHz, from the combined data of MERLIN and the VLA, is
genuinely associated with the sub-mmm source \cite{dunlop02}.
Furthermore this strengthens the suggestion that a marginal
3.3$\sigma$ detection at 8.4\,GHz (VLA 3651+1226, Richards,
priv. comm.) that may also associated with HDF850.1.

The redshift probability distribution for HDF850.1, based on the
multi-wavelength 170$\mu$m--1.4\,GHz data, indicates 
$z$ is comprised between 3.3 and 5.4
with a 86\% to 90\% confidence depending on the adopted prior 
(Fig.\ref{fig:hdf_co}).  This high
photometric redshift supports the earlier decision to reject the $z\sim 1$
elliptical galaxy 3$-$586, as the optical counterpart. Unfortunately
the IR counterpart, HDF850.1K, is too faint to attempt IR spectroscopy
with even the 10-m class of telescopes.

The most efficient means to determine the spectroscopic redshifts of
optically-obscured sub-mm galaxies is the detection of millimetre and
centimetre wavelength molecular lines using large single-dish
telescopes (e.g. GBT or LMT) with broad-band receivers (Townsend et
al. 2001, 
Carilli \& Blain 2002). In this way, it
is possible to eliminate the necessity for an optical/IR counterpart,
or a position accurate to $< 3$\,arcsec.  
Given the predicted accuracy of the photometric redshift for HDF850.1, 
and taking all of the uncertainties discussed in section~2 into
account, we demonstrate in Fig.\ref{fig:hdf_co} that the J=1--0 CO line
in HDF850.1 has a 86-90\% probability of detection in the K-band
receiver (18.0--26.5 GHz) on the GBT.
Observations to conduct searches for CO emission in high-$z$ sub-mm
galaxies, guided only by their photometric redshifts, are currently
underway on the GBT.

\section{Conclusions}

We have derived meaningful photometric-redshift probability
distributions for individual sub-mm galaxies using Monte Carlo
simulations that take into account a range of measurement errors and
model-dependent uncertainties. From the individual redshift
probability distributions, we have measured the cumulative redshift
distribution for the sub-mm population of galaxies identified in blank
field surveys.  We find that the multi-wavelength FIR--radio data for
50 sub-mm galaxies detected ($> 3.5\sigma$) at 850$\mu$m are
consistent with $\sim 50-90$\% of the population lying at redshifts $z
= 2- 4$, with only a small fraction, $\sim 10$\% of the population at
$z < 2$. We also show that up to 50\% of the galaxies have colours
that are consistent with $z > 4$. The possibility that the sub-mm
galaxy population contains a significant fraction 
undergoing a major burst of star formation at such early epochs has
important consequences for models of galaxy formation. This suggestion 
is still highly dependent on the prior assumption of the evolutionary
model for the
  sub-mm population which affects most strongly those 
  sub-mm sources 
  detected only at $850\mu$m.

Shorter-wavelength sub-mm data ($250-500\mu$m) from a
future balloon-borne experiment, BLAST, will provide powerful additional
constraints ($\Delta z \sim \pm 0.5$) on the redshift distribution of
all the SCUBA and MAMBO galaxies at $z > 2$. A photometric-redshift
accuracy of $\pm 0.5$ for an individual galaxy will be sufficient to
provide robust statistical measurements of the redshift distribution
of the entire population, the global history of dust-obscured star formation
and the clustering properties of sub-mm galaxies.

We have argued that the
large dispersion in the radio to sub-mm SEDs (or 850$\mu$m/1.4GHz
colour) of the template galaxies naturally translates into an
imprecise measure of the photometric redshift, and that radio
observations cannot be used efficiently to measure the individual
redshifts of thousands of galaxies detected in future wide-area sub-mm
surveys. The sub-mm/radio spectral index is still able to 
discriminate between the $z<2$ and $z>2$ regime.

We have an on-going programme to conduct centimetre-wavelength
observations of CO J=1-0 molecular line transitions in high-$z$ sub-mm
galaxies on the 100-m GBT. The observations, however, first require an
accurate photometric-redshift probability distribution to guide the
choice of receiver and tuning parameters.  We illustrate the power of
our photometric-redshift technique with the example of HDF850.1, one
of the best-studied sub-mm galaxies \cite{hughes98,downes99,dunlop02}.

Within the next few years, the combination of ground-based,
balloon-borne and satellite experiments will provide accurately
calibrated rest-frame sub-mm--FIR data for a significant population of
sub-mm galaxies. The centimetre and millimetre wavelength
spectroscopic redshift confirmation of their photometric redshifts
will calibrate the method outlined in this paper. Thus we can finally
look forward to breaking the ``redshift deadlock'' that
currently prevents an accurate understanding of the nature of the
sub-mm galaxy population and their evolutionary
history.

\section{Acknowledgments}
This work has been partly supported by CONACYT grants 32180-E and 32143-E.
We thank the anonymous referee for the critical comments and careful reading
of this manuscript.

%\end{document}

\appendix

\section{Redshift estimates for individual sub-mm galaxies}

A catalogue of redshift probability distributions and SEDs for the
galaxies detected 
in more than 2 bands (Tables~1 and 2) is included. We also include the
corresponding colour-colour-redshift or colour-flux-redshift plots from 
which they have been derived. All estimates are derived for model le2.

For galaxies detected in just one band (Table~3) we present a selection
of the results derived for 6 out of the 46 galaxies included in Table~3.
The general uniformity of their redshift distributions is a fair 
representation of those galaxies not included in this appendix.

\end{document}
\include{figapp3}

\include{figapp2}

\newpage
\begin{figure*}
    \vspace*{-3.5cm}
    \hspace*{14cm}
    \figl{12cm}{26}{26}{573}{737}{d21c.ps}{90} 

    \caption{Selection of flux-redshift plots, redshift distributions and 
comparison of the template SEDs and observed data 
for sub-mm sources detected in 1 band at a $>3\sigma$ level.
The slanted box in the flux-redshift plot marks the $1\sigma$ error box of 
the sub-mm galaxy. For the sake of clarity, 
just 1000 galaxies from the mock catalogue that fulfill the observational
requirements of the sub-mm galaxies
are represented (as dots). 
The rest of the elements in the panels are as in Fig.~A1,  except for the
template SEDs which are compatible the colours of the sub-mm galaxies 
at a $3\sigma$ level, which are represented in 
dark grey here.
}
\end{figure*}

\newpage
\begin{figure*}
 \setcounter{figure}{2}                                                       

    \vspace*{-3.5cm}
    \hspace*{14cm}
    \figl{12cm}{26}{26}{573}{737}{d13c.ps}{90} 

    \vspace*{-4cm}
    \hspace*{14cm}
    \figl{12cm}{26}{26}{573}{737}{d2c.ps}{90} 
    \caption{(cont.)}
\end{figure*}

\newpage
\begin{figure*}
 \setcounter{figure}{2}                                                       
    \vspace*{-3.5cm}
    \hspace*{14cm}
    \figl{12cm}{26}{26}{573}{737}{d6c.ps}{90} 

    \vspace*{-4cm}
    \hspace*{14cm}
    \figl{12cm}{26}{26}{573}{737}{d83c.ps}{90} 
    \caption{(cont.)}
\end{figure*}

\newpage
\begin{figure*}
 \setcounter{figure}{2}                                                       

    \vspace*{-3.5cm}
    \hspace*{14cm}
    \figl{12cm}{26}{26}{573}{737}{d5c.ps}{90} 
    \caption{(cont.)}
\end{figure*}

%\section{SEDs}         
%
%Here we include the photometry of the template SEDs used in the analysis 
%of photometric redshifts.
%
%
%\begin{figure*}
%\vspace{7.0cm}
%\special{psfile=sedIIpaper.ps angle=90 hscale=38 vscale=38
%hoffset=280 voffset=-20}
%\special{psfile=sedIIpaperb.ps angle=90 hscale=38 vscale=38
%hoffset=530 voffset=-20}
%\caption{{\it Left panel:} Rest-frame spectral energy distributions
%(SEDs) of 20 starburst galaxies, ULIRGs and AGN, normalized at
%60$\mu$m.  Lines represent the best fit SED models to the data
%and include contributions from non-thermal synchrotron
%emission, free-free and grey-body thermal emission.; 
%{\it Right panel:} Composite SED
%models for the identical 20 template galaxies shown in the left panel,
%normalized at 60$\mu$m and scaled to the flux density of Arp220 at
%each wavelength.  The vertically bounded-regions (slanted and crossed) 
%correspond to the rest-frame wavelength regimes covered by
%observations at 1.4~GHz and 850$\mu$m filters
%respectively for galaxies at $z=0-6$.  This
%representation of the SEDs shows more clearly the wavelength regimes
%in which the minimum dispersion in galaxy colours can be
%expected. }
%\label{fig:seds1}
%\end{figure*}

\end{document}